%% file: arXiv_LiDAR_Wireless.tex
\documentclass[journal,twocolumn]{IEEEtran}
\input{input.tex}

\usepackage[table]{xcolor}
\usepackage{epstopdf}
\usepackage{makecell}
\usepackage{array,colortbl,multirow,multicol,booktabs,ctable}
\usepackage{enumerate}
\usepackage{algorithmicx}
\usepackage{algorithm}
\usepackage{amsmath}
\usepackage[noend]{algpseudocode}
\usepackage{float}
\usepackage{soul}
\usepackage{hyperref}
\usepackage{color}
\usepackage{makeidx}
\usepackage{bbm}
\usepackage{graphicx}
\usepackage{caption}
\usepackage{array}
\usepackage{epsfig}
\usepackage{comment}
\usepackage{textcomp}
\usepackage{subcaption}
\usepackage{cite}
\usepackage{balance}
\usepackage[font=small]{caption}

\newcommand{\sref}[1]{{Section}~\ref{#1}}

\newcommand{\pinv}[1]{\ensuremath{#1^{\dagger}}} 	

\begin{document}
	\title{Proactively Predicting Dynamic 6G Link Blockages Using LiDAR and In-Band Signatures}

	\author{Shunyao Wu, Chaitali Chakrabarti, and Ahmed Alkhateeb\\  \thanks{The authors are with the School of Electrical, Computer and Energy Engineering, Arizona State University, (Email: vincentw, chaitali, alkhateeb@asu.edu). This work was supported in part by the National Science Foundation under Grant 2048021. Part of this work is accepted by the IEEE Wireless Communications and Networking Conference (WCNC) 2022 \cite{wu2021lidar}.}}
	\maketitle

\begin{abstract}
Line-of-sight link blockages represent a key challenge for the reliability and latency of millimeter wave (mmWave) and terahertz (THz) communication networks. To address this challenge, this paper leverages mmWave and LiDAR sensory data to provide awareness about the communication environment and proactively predict dynamic link blockages before they occur. This allows the network to make proactive decisions for hand-off/beam switching, enhancing the network reliability and latency. More specifically, this paper addresses the following key questions: (i) Can we predict a line-of-sight link blockage, before it happens, using in-band mmWave/THz signal and LiDAR sensing data? (ii) Can we also predict when this blockage will occur? (iii) Can we predict the blockage duration? And (iv) can we predict the direction of the moving blockage? 
For that, we develop machine learning solutions that learn special patterns of the received signal and sensory data, which we call \textit{pre-blockage signatures}, to infer future blockages. To evaluate the proposed approaches, we build a large-scale real-world dataset that comprises co-existing LiDAR and mmWave communication measurements in outdoor vehicular scenarios. Then, we develop an efficient LiDAR data denoising algorithm that applies some pre-processing to the LiDAR data.  Based on the real-world dataset, the developed approaches are shown to achieve above 95\% accuracy in predicting blockages occurring within 100 ms and more than 80\% prediction accuracy for blockages occurring within one second. Given this future blockage prediction capability, the paper also shows that the developed solutions can achieve an order of magnitude saving in network latency, which further highlights the potential of the developed blockage prediction solutions for wireless networks.   
\end{abstract}



\maketitle

\section{INTRODUCTION} \label{sec:Intro}
e the high data rate gains of millimeter wave (mmWave) and sub-terahertz (sub-THz) communications requires overcoming a number of key challenges \cite{Rappaport2019,Rappaport2013a}. One of these important challenges is the Line of Sight (LoS) link blockages which could cause sudden link failures, impacting the reliability and latency of the mobile networks. This is particularly important at mmWave/sub-THz systems because of (i) their reliance on LoS communications for sufficient receive signal power and (ii) the high penetration loss of these high-frequency signals (high sensitivity to blockages). One promising approach for addressing LoS link blockage challenge is by \textit{proactively} predicting link blockages before they happen \cite{alkhateeb2018machine, mmwave_journal}, thereby allowing the network to make proactive beam switching/basestation hand-off decisions. This proactive prediction method relies on the use of machine learning models that can potentially utilize side information and prior observations to predict future blockages \cite{alkhateeb2018machine, Alrabeiah2020a,mmwave_journal}. In our previous work \cite{mmwave_journal}, we proactively predicted the blockage with high accuracy when the prediction interval was short ($\sim$ 300 ms). To proactively predict the blockage for longer time intervals, additional data modality such as LiDAR is needed. In \cite{wu2021lidar}, we showed that LiDAR data can help achieve high accuracy for longer prediction intervals. In this paper, \textbf{we investigate the potential of leveraging \textit{LiDAR and wireless mmWave/sub-THz signatures} to proactively predict future link blockages}. In particular, we attempt to answer the following four questions: (i) Can mmwave/THz signal and LiDAR sensing data be utilized to predict future blockages? (ii) Can these signals also predict when a blockage will happen in the future? (iii) Can these signals be used to predict the type of the blockage? And (iv) can these in-band wireless signals be leveraged for blockage direction prediction?

\subsection{Related Work}
Initial solutions for the mmWave blockage challenges focused on multiple-connectivity where a user simultaneously maintains multiple links with multiple base stations \cite{giordani2016multi,polese2017improved,petrov2017dynamic,aziz2016architecture}. For instance,  \cite{giordani2016multi} developed a multiple-connectivity scheme that enables efficient and highly adaptive cell selection in the presence of mmWave channel variability. In \cite{polese2017improved}, an uplink control signaling system combined with a local coordinator is proposed for multiple-connectivity, which enables rapid path switching. A methodology that combines concepts from queuing theory and stochastic geometry for dynamic multiple-connectivity is presented in \cite{petrov2017dynamic}. In this integrated framework,  both user- and network-centric performance indicators were quantified to highlight the effectiveness of the proposed framework in representative mmWave scenarios. Further, mmWave systems assisted by sub-6GHz links are developed in \cite{Mismar2021,aziz2016architecture} to realize fast link switching and efficient multiple-connectivity. While multiple-connectivity is a promising approach for enhancing the mmWave network reliability, it has several drawbacks: (i) These approaches require maintaining multiple simultaneous links which consume and under-utilize the network resources. For example, in the multiple connectivity case, if two or more basestations have active link connections with the user, then two or more times of the network resources are unnecessarily allocated or reserved. In our proposed solution, only one basestation (the serving basestation) needs to keep a connection with the user, as it can proactively hand-off the user to another basestation if a future blockage is detected. This saves half the network resources (as the other basestation will not dedicate resources until the proactive hand-off process is done). (ii) These multiple-connectivity solutions \textit{react} to the blockages after they happen and thus they still incur high network latency and do not completely address the reliability and latency challenges associated with LoS link blockages in mmWave and THz systems. 

Machine learning (ML) models were successfully used to proactively predict future blockages in \cite{alkhateeb2018machine,alrabeiah2020deep,Ali2020,charan2021vision,Charan22a,mmwave_journal,Demirhan_mgazine_radar}. In particular, the authors in \cite{alkhateeb2018machine} developed a recurrent neural network that used the sequence of beams of the previous few instances to proactively predict future LoS link blockages. The solution in \cite{alkhateeb2018machine}, however, was targeted for stationary blockages. To predict future dynamic/moving blockages, the authors in \cite{alrabeiah2020deep,Ali2020} proposed to make use of the sub-6GHz channels to infer link blockages in near future. Further, the approach in \cite{mmwave_journal} achieved a high prediction accuracy using in-band mmWave data when the prediction time interval was small; the system did not perform as well when the prediction time interval became longer. This motivated looking into the use of multiple data modalities for future blockage prediction. The method in \cite{charan2021vision,Charan22a} leveraged the rich environment sensory data obtained, for example, from visual sensors (cameras) to proactively predict link blockages well before they happen. Deploying cameras, however, may not always be possible for privacy/regulatory reasons. This prompted the research for using other data modalities for future blockage prediction. With this objective, in  \cite{wu2021lidar}, we proposed for the first time to use LiDAR sensory data for proactively predicting link blockages. Motivated by the initial results in \cite{wu2021lidar}, this paper extends this work and explores the potential of utilizing \textit{both} LiDAR and in-band mmWave measurements to predict the occurrence, severity, and direction of dynamic link blockages. 

Note that a relevant line of work explores how ML could be leveraged to estimate/predict the mmWave channels or their parameters (angles or arrival/departure, delay, etc.) \cite{Moon2020,alrabeiah2020deep,Wei2021,Ma2021,Huang2018}. The channel estimation line of work does not address the blockage prediction questions. For example, it does not proactively predict if a blockage will happen in the future and when it will happen. It is also important to differentiate between future blockage prediction and current blockage status prediction/identification (also called LoS/NLoS link identification) which has also been studied in recent papers \cite{choi2017deep,huang2020machine,Alrabeiah2020a}. For example, the work in \cite{choi2017deep,huang2020machine} leveraged machine/deep learning to identify whether the current receive signal corresponds to a LoS or NLoS link. Further, \cite{Alrabeiah2020a} used a camera feed along with sub-6 GHz channels to identify currently blocked mmWave links. While identifying the current link status has important applications, our focus in this paper is on proactive/future blockage prediction, which could bring promising reliability and latency gains for mmWave and sub-terahertz networks. 

\begin{figure*}[t]
	\centering
	\includegraphics[width=0.8\linewidth]{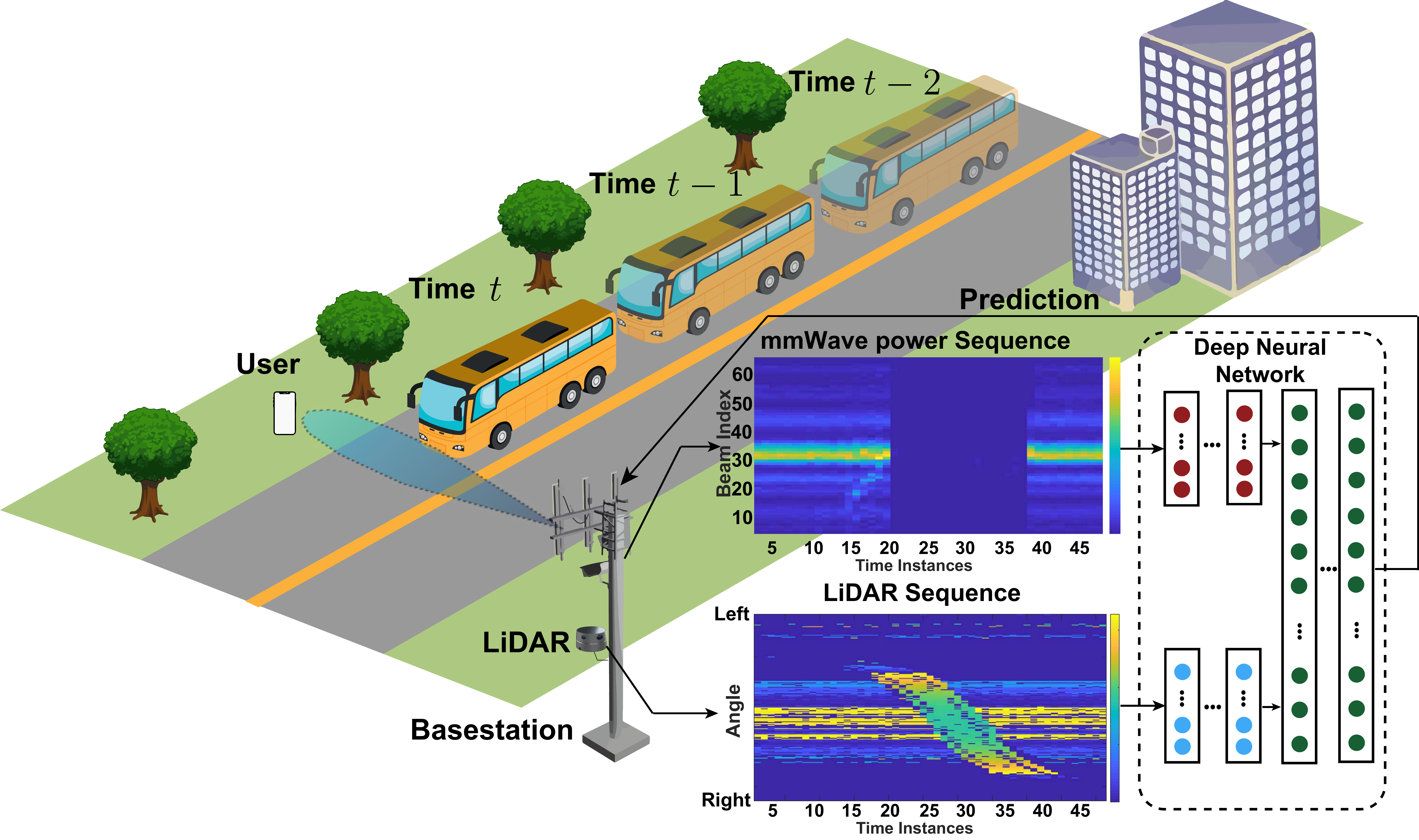}
	\caption{Overall system model where a mmWave/sub-THz basestation leverages both LiDAR and in-band wireless sensing to provide environment awareness and enable the proposed proactive line-of-sight link blockage prediction approach.}
	\label{fig:system diagram}
\end{figure*}

\subsection{Contribution} 
In this paper, we propose to leverage both LiDAR sensory data and mmWave data to provide information about the communication environment and help proactively predict LoS mmWave link blockages potentially seconds before they happen. The main contributions of the paper can be summarized as follows:
\begin{itemize}

	\item We propose a new approach that leverages fused mmWave signals and LiDAR sensory data to proactively predict future dynamic LoS link blockages before they occur. This approach achieves high accuracy (95\%) for blockage prediction and can work in conjunction with regular communication systems. 
	
	\item We develop baseline LiDAR methods and ML models based on convolutional neural networks that utilize receive mmWave signals and/or LiDAR sensory data to address four ML tasks: (i) Predicting the occurrence of future blockages, (ii) predicting when, in the future, will the blockage occur, (iii) classifying the type of blockage, and (iv) classifying the moving blockage direction. A trade-off discussion of the performance, complexity, and universality between using one source and fused sources is conducted for these four problems.
	
	\item We construct a framework for collecting co-existing mmWave and LiDAR data and use it to build a large-scale real-world outdoor dataset ($\sim$ 500 thousand data points). The dataset is collected at three different locations and consists of mmWave beam training data, LiDAR sensory data, and the corresponding images, and can be used to study multiple problems including the LiDAR-aided blockage prediction task. 
	
	\item A static cluster removal (pre-processing) algorithm for the data generated by low-cost LiDAR sensors is proposed. The approach denoises this data and accurately extracts the traces of the moving objects, thus improving the performance of our ML models.
\end{itemize}

Using the developed dataset, we validate the feasibility of the proposed proactive blockage prediction approach and draw important insights into its performance in outdoor wireless communication scenarios. Overall, the results highlight the potential of using the developed solution in accurately predicting future blockages, which could enhance the reliability and latency performance of future mmWave and sub-THz networks.

\subsection{Organization}
The rest of the paper is organized as follows: The system and channel models are introduced in \sref{sec:sys_model}. \sref{sec:prob_formul} defines and formulates the four blockage prediction problems. The proposed baseline LiDAR method and ML-based blockage prediction approaches are then presented in \sref{sec:key_idea}, \sref{sec:non-ML_method}, and \sref{sec:ML_method}. To evaluate the proposed approaches, we built a hardware testbed and collected a large-scale dataset that is described in \sref{sec:exp_testbed} and dataset pre-processing in \sref{sec:dev_dataset}. Finally, the experimental results for evaluating the proposed blockage prediction solutions are presented in \sref{sec:results}. 

\textbf{Notation}: We use the following notation throughout this paper: $\bA$ is a matrix, $\ba$ is a vector, $a$ is a scalar, and $\cA$ is a set. $|\bA|$ is the determinant of $\bA$, $\|\bA \|_F$ is its Frobenius norm, whereas $\bA^T$, $\bA^H$, $\bA^*$, $\bA^{-1}$, $\pinv{\bA}$ are its transpose, Hermitian (conjugate transpose), conjugate, inverse, and pseudo-inverse respectively. $[\bA]_{r,:}$ and $[\bA]_{:,c}$ are the $r$th row and $c$th column of the matrix $\bA$, respectively. $\mathrm{diag}(\ba)$ is a diagonal matrix with the entries of $\ba$ on its diagonal. $\bI$ is the identity matrix and $\mathbf{1}_{N}$ is the $N$-dimensional all-ones vector. $\bA \otimes \bB$ is the Kronecker product of $\bA$ and $\bB$, and $\bA \circ \bB$ is their Khatri-Rao product. $\cN(\bm,\bR)$ is a complex Gaussian random vector with mean $\bm$ and covariance $\bR$. $\bbE\left[\cdot\right]$ is used to denote expectation.

\section{System Model} \label{sec:sys_model}
We adopt a mmWave communication system where a basestation with $M_\mathrm{A}$-element antenna array is used to serve a stationary user. The basestation is further equipped with a LiDAR sensor to provide awareness about the surrounding environment and moving scatterers/blockages, as shown in \figref{fig:system diagram}. The basestation employs a pre-defined beamsteering codebook of $M$ beams, $\boldsymbol{\mathcal F} = \{\mathbf f_m\}_{m=1}^{M}$, where $\mathbf f_m$ is a beamsteering vector that directs the signal towards direction $\theta_m= \theta_\text{offset}+ \text{FoV}/M$, with FoV denoting the field of view of the wireless beamforming system \cite{Zhang_RL}. In our testbed, described in \sref{sec:exp_testbed}, we consider a phased array with $M_\mathrm{A}$=16 elements and a  codebook of $M=64$ beamforming vectors, with steering angles uniformly quantized in the range $[-\pi/4, \pi/4]$. It is worth mentioning here, though, that the proposed blockage prediction approaches in this paper can be applied to more general array architectures.

To account for the variations of the channel over time, we adopt a block fading channel model where the channel is assumed to be constant over a time duration of $\tau_B$. Further, we adopt an OFDM signal transmission model of $K$ subcarriers. We define $\bh_k[t] \in \mathbb C^{M\times 1}$ as the downlink channel from the base station to the user at the $k$-th subcarrier for discrete time instance $t$, where $t\in \mathbb Z$.  At time $t$, if the beamsteering vector $\bff_m$ is adopted by the basestation for the downlink transmission, then the received signal at subcarrier $k$ is 
\begin{equation}\label{sig_model}
r_{k,m}[t] = \bh_k[t]^T \bff_m s_k[t] + n_k[t],
\end{equation}
where $s_k[t]$ is the transmitted symbol at the $k$-th subcarrier and $t$-th time instance, $\bbE{\left|s_k[t]\right|^2}=1$, and $n_k[t] \sim \mathcal {CN}(0,\sigma_n^2)$ is a noise sample.

\section{Problem Formulation} 
\label{sec:prob_formul}
Proactively identifying the Line of Sight (LoS) link status significantly benefits both the physical and the network layers. In this paper, we propose the following four problems and address them. (i) How to leverage the received mmWave signal power and LiDAR sensory information to predict the occurrence of the blockage in the near future or not? (ii) In case there is a blockage, how to use the received mmWave signal power information and LiDAR sensory data to predict the exact time that blockage will occur? (iii) Given the knowledge of an incoming blockage, is it possible to predict its size (link blockage time interval)? and, finally, (iv) Could the direction of the blockage be predicted? A more formal description of each problem is included below

\textbf{Problem 1. Blockage Occurrence Prediction:} The first problem aims to proactively predict the occurrence of a blockage in the near future. This problem is formulated as follows. Define $t\in \mathbb Z$ as the index of the discrete time instance, and $x[t]\in{0, 1}$ as the link status (blocked or unblocked) at the $t$th time instance. $x[t]=1$ indicates a blocked link, i.e., it means that the LoS path between the transmitter and receiver is blocked, while $x[t]=0$ indicates an unblocked link. Further, if at time instance $t$, the user transmits a pilot signal to the base station and the base station receives this signal using $M$ beams in the codebook $\boldsymbol{\mathcal F}$, then the receive power vector of the $M$ beams at the $t$th instance is defined as
\begin{equation} \label{eq:rec}
	\mathbf r[t] = \left[ |r_{1}[t]|^2, \dots, |r_{M}[t]|^2\right]^T.
\end{equation}
Note that $|r_{m}[t]|^2$ is the total power over the $K$ subcarriers, i.e., $|r_{m}[t]|^2=\sum_{k=1}^K |r_{k,m}[t]|^2$. The sequence of these receive power vectors for the $T_{ob}$ previous time instances (observation window), $t-T_{ob}+1$, ... $t$, is combined to form another sequence, $\mathcal{R}_{ob}$, which is expressed as
\begin{equation}
	\mathcal{R}_{ob} = \{ \mathbf r[t+n] \}_{n=-T_{ob}+1}^{0}.
\end{equation} 

We assume that at time instance $t$, the LiDAR sensor provides sensory data $\mathbf L[t] \in \mathbb{R}^{P \times 2}$, where $P$ represents the number of points collected by the LiDAR sensor at time $t$ for a 360-degree scan; each data point consists of an angle and a distance value representing the measured depth at this angle. We define $\mathcal{L}_{ob} $ as the observation sequence of LiDAR samples at the $T_{ob}$ previous time instances (observation window):  
\begin{equation}
\mathcal{L}_{ob} = \{ \mathbf L[t+n] \}_{n=-T_{ob}+1}^{0}.  
\end{equation}
Then our observation sequence is expressed as 
\begin{equation}
\mathcal{S}_{ob} = \{ \mathcal{R}_{ob},\mathcal{L}_{ob} \}.  
\end{equation}
\noindent Now, given the observed sequence $\mathcal S_{ob}$, our goal is to predict the occurrence of link blockage within a future time interval ($T_P$ instances). We use $b_{T_P}$ to indicate whether there is a blockage occurrence within that future interval or not. More formally, $b_{T_p}$ is defined as follows
\begin{equation} \label{equ:p1_label}
	b_{T_P} = 
	\begin{cases}
		0, &  x[t+n_{p}] = 0 \quad \forall n_{p} \in \{ 1,\dots,T_P \} \\
		1, & \text{otherwise},
	\end{cases}       
\end{equation}
where $1$ indicates the occurrence of a blockage and $0$ is the absence of a blockage. The goal of this problem is then to predict $b_{T_p}$ with high accuracy. If $\hat {b}_{T_p}$ denotes the predicted link status, {the objective of Problem 1} is then to maximize the prediction success probability, given an observed sequence of receive power and LiDAR sensory points $S_{ob}$ i.e., 
\begin{equation} \label{equ:p1_object}
	\textbf{Objective 1:} \ \ \text{Maximize} \ \mathbb P_1( \hat {b}_{T_p}= b_{T_p} | \mathcal S_{ob}),
\end{equation}

\begin{figure}[t]
	\centering
	\includegraphics[width=\columnwidth]{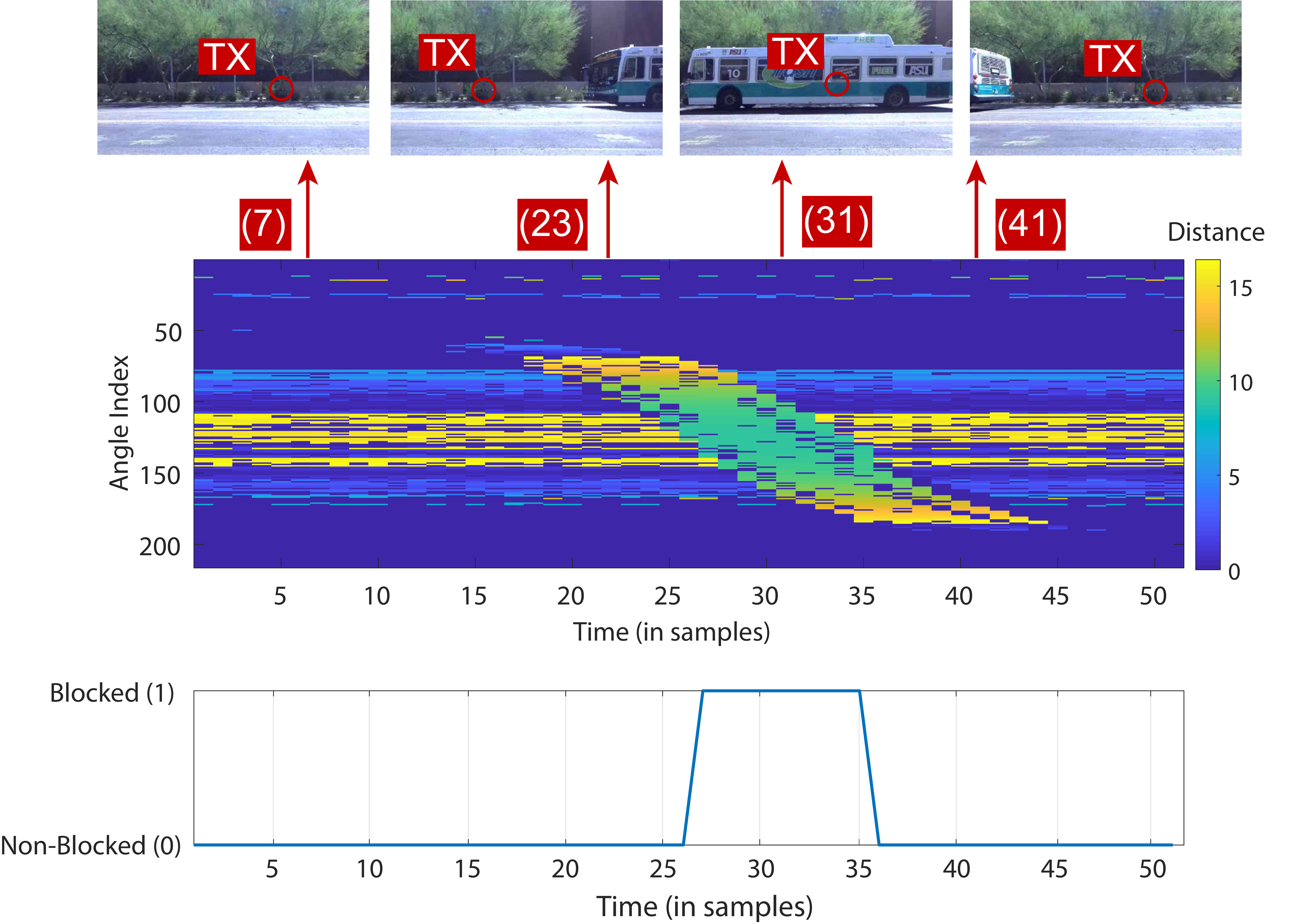}
	\caption{An example of outdoor pre-blockage signature: The upper images show the real scene at the location we collected the data. These images are taken by a camera deployed right below the received phased array. The TX is identified in the red circle. The middle figure shows the distance heatmap for every angle versus time. The bottom figure illustrates the corresponding link status.}
	\label{fig:lidar_exp}
\end{figure}

\textbf{Problem 2. Blockage-instance prediction:} Given the observed sequence $\mathcal S_{ob}$ and the knowledge that there is a blockage occurring within the future $T_p$ instances, the second problem aims to identify the time instance when the blockage would occur. In particular, we want to predict the exact time instance $n_{p}$ at which $x[t+n_{p}] = 1$. Given the observed sequence $\mathcal S_{ob}$ due to an incoming blockage, the objective of Problem 2 is to minimize the absolute error between the predicted time instance of blockage occurrence, $\hat{n}_{p}$, and the exact time instance of blockage occurrence $n_p$.
\begin{equation} \label{equ:p2_object}
	\textbf{Objective 2:} \ \ \text{Minimize} \ \mathbb E_2(abs(\hat{n}_{p} - n_{p})|\mathcal S_{ob}, b_{T_p}=1).
\end{equation}

\textbf{Problem 3. Blockage-severity prediction:} Moving objects in the wireless environment take different shapes, sizes, and speeds, and, hence, the duration of link blockages varies. Problem 3 focuses on proactively identifying the type of the incoming blockage from the perspective of its \textit{severity}. We define \textit{blockage-severity index} as the average blockage time interval measured in a number of instances (or blocks $\tau_B$) for each object\footnote{Note that this average duration is a function in different aspects of the object such as speed, shape, and size.}. We classify objects based on this index. 

Let $\mathcal V$ define the set of all physical objects that could be present in the environment, e.g., car, bus, tree,... etc\footnote{These objects are labeled based only on discernible visual traits, something similar to that in \cite{ImageNet}}, and let $\bar{\rho}_v$ define the average blockage time interval of the object $v\in\mathcal V$. The blockage-severity index for object $v$ is defined as
\begin{equation} \label{equ:p3_label}
	b_{\text{sev},v} = \left\{ \begin{array}{ll}
		1, \quad & \hat{\rho}_v \in T_{D_1},\\
		2, \quad & \hat{\rho}_v \in T_{D_2}, \\
		\vdots & \\
		N_\mathrm{class}, \quad & \hat{\rho}_v \in T_{D_{N_\mathrm{class}}}, \\
	\end{array} \right.
\end{equation}
where $N_\mathrm{class}$ is the total number of classes based on the length of time interval the object blocks the LoS link, and $T_{D_1},\dots,T_{D_{N_\mathrm{class}}}$ are consecutive time intervals measured in terms of time instances. Those intervals and their number of classes $N_\mathrm{class}$ are determined by partitioning set $\{\rho_v\}_{v=1}^{|\mathcal V|}$ into $N_\mathrm{class}$ partitions using a distortion metric. For example, the average  blockage time interval ($\bar{\rho}_v$) of walking humans is 237 ms, while 347 ms for sedans and 396 ms for SUVs. So if $T_{D_1}$ is 0-300 ms and $T_{D_2}$ is 300 ms-600 ms, then walking human belongs to severity level 1, sedans and SUVs belong to severity level 2. More on the partitioning process in a certain wireless environment is presented in \cite{mmwave_journal}. Now, if $\hat{b}_{\text{sev},v}$ denotes the predicted blockage severity index, the objective of Problem 3 is to maximize the probability of successful severity level prediction given the condition that the blockage will occur.
\begin{equation} \label{equ:p3_object}
	\textbf{Objective 3:} \ \ \text{Maximize} \ \mathbb P_3(\hat{b}_{\text{sev},v} = {b}_{\text{sev},v}|\mathcal S_{ob}, b_{T_p}=1).
\end{equation}  

\textbf{Problem 4. Blockage-direction prediction:} Another interesting dimension of proactive blockage prediction is predicting the direction of moving blockage. Let $b_{\text{dir}}\in\mathcal D$ be a variable indicating the motion direction of a blocking object where the motion direction is a value from a finite set of possible pre-defined directions $\mathcal D = \{0,1,\dots,G\}$. In this paper since we only consider the transmitter and receiver to be positioned on two opposite sidewalks of a city street, there are only 2 possible directions $\mathcal D = \{0,1\}$, where 0 indicates a vehicle traveling left to right with respect to the receiver and 1 indicates a vehicle towards right to left direction. The objective of Problem 4 is to maximize the probability of successful prediction of moving direction given the condition that the blockage will occur: 
\begin{equation} \label{equ:p4_object}
	\textbf{Objective 4:} \ \ \text{Maximize} \ \mathbb P_4( \hat {b}_{\text{dir}} = b_{\text{dir}}| \mathcal S_{ob}, b_{T_p}=1).
\end{equation}  
where $\hat b_{\text{dir}}$ is the predicted motion direction and $b_{\text{dir}}$ is the groundtruth motion direction of the object.

All four problems are inherently proactive, and this calls for the ability to sense the environment and identify patterns that characterize blocking objects and their behaviors. We propose a baseline LiDAR method and an ML-based approach  to tackle these problems. For the ML-based approach, the algorithm is trained to identify patterns in $\mathcal S_{ob}$ and use them to perform a task of interest. Formally, the proposed approach could be described as a function learning problem with a training dataset. More specifically, a function $f_{\Theta} (\mathcal S_{ob})$ is learned by training an ML algorithm to estimate a set of parameters $\Theta$ from a training dataset. In this paper, we use $f_{\Theta_1}$, $f_{\Theta_2}$, $f_{\Theta_3}$, $f_{\Theta_4}$ for {Problems 1, 2, 3, 4}, respectively. The training is carried out to maximize the corresponding success probability, i.e., $\mathbb P_1(.)$, $\mathbb P_2(.)$, $\mathbb P_3(.)$, and $\mathbb P_4(.)$ defined in equations \ref{equ:p1_object}, \ref{equ:p2_object}, \ref{equ:p3_object}, \ref{equ:p4_object}.

\section{Proactive Blockage Prediction using Wireless and LiDAR Signatures} \label{sec:key_idea}
In our previous work \cite{mmwave_journal}, we showed that leveraging wireless pre-blockage signatures could be a promising way for proactive blockage prediction. The signatures based method is highly effective in predicting blockages that are very close to occurrences, i.e. within a very short window ~ 100-200ms. In this work, we leverage LiDAR sensory data that can work independently or jointly with the wireless signatures for future blockage prediction. 

A typical LiDAR sensor sends pulsed light waves into the surrounding environment. These pulses are reflected by the objects and returned to the sensor, and the sensor uses the round-trip time to calculate the distance it traveled. By sending and receiving laser beams, the LiDAR sensor collects a 2-D point cloud map of the surrounding environment. We propose to leverage these LiDAR sensory data to detect if an object is going to block the mmWave communication link between the basestation and the mobile user. \figref{fig:lidar_exp} shows an example when a communication link is getting blocked by a moving object. At the top panel, the images (with TX labeled in red circle) represent the real scenes at RX perspective when the bus blocked the communication link. The heatmap in the middle represents the sensed distance for every quantized angle (direction) level as a function of time. The horizontal lines represent static objects (since their distances from the LiDAR device do not change over time). The bottom plot shows the link status and the bottom photos show the corresponding scenario. Note that from time instances 13 to 26, the color of the pattern changes from yellow to green, indicating that the distance between the moving object (bus) and the LiDAR sensor has become shorter. This matches the scenario of an approaching bus. The color becomes yellow again after time instance 35 implying that the object has moved away from the LiDAR sensor. \textbf{Thus, as the blockage approaches the link, we can see a clear pre-blockage pattern in the LiDAR heatmap that can potentially be leveraged for proactive blockage prediction.} In the next section, we focus on using LiDAR sensory data and propose LiDAR-only baseline solutions for proactive blockage prediction. Then, in Section \ref{sec:ML_method}, we present the ML approaches that proactively predict future blockages using \textit{both} LiDAR and wireless signatures.

\section{Proposed Baseline LiDAR Method} 
\label{sec:non-ML_method}
Recall that \textbf{Problem 1} predicts the occurrence of blockage, \textbf{Problem 2} predicts the exact time of blockage occurrence, \textbf{Problem 3} predicts the severity level of the blockage and \textbf{Problem 4} predicts the motion direction of the blockage. A threshold-based approach for \textbf{Problem 1} and \textbf{Problem 3} is proposed in \sref{subsec:non-ML_p1} and \sref{subsec:non-ML_p3}, and a least square estimation method for \textbf{Problem 2} is presented in \sref{subsec:non-ML_p2}. An average location comparison method is used for \textbf{Problem 4}. \textbf{All non-ML-based methods are based on processing only LiDAR data.}

\begin{figure*}[tb]
\centering
\includegraphics[width=\linewidth]{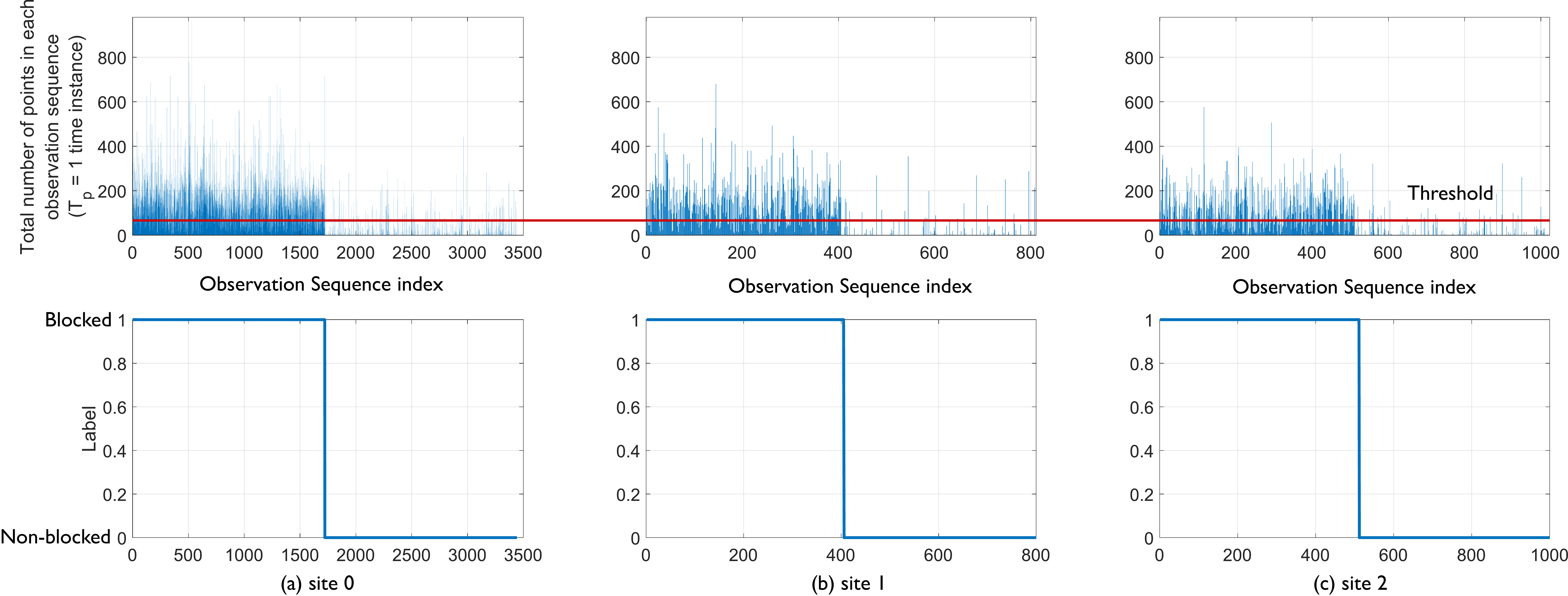}
\caption{Distribution of the number of LiDAR points for each observation sequence in the first time instance in (a) site 0, (b) site 1 and (c) site 2. We calculate the total number of LiDAR points in each observation sequence when the prediction interval is 1 time instance.}
\label{fig:p1_dis}
\end{figure*}

\subsection{Blockage Occurrence Prediction (\textbf{Problem 1})} \label{subsec:non-ML_p1}
\textbf{Problem 1} focuses on predicting the occurrence of blockage. The difference between the LiDAR point cloud when a blockage exists or does not is some extra points that show the trace of the incoming blockage. Thus, the total number of points in a LiDAR observation sequence could potentially be utilized to determine if a blockage is going to block the link. In our baseline solution, we use a threshold on the number of LiDAR cloud points at time instance $t$ and angle $\Theta$ to determine the existence of a possible blockage. 
\begin{equation} \label{equ:p1_non_ML}
    \hat {b}_{T_p} = 
    \begin{cases}
      1, &   \sum_{t=1}^{T_{ob}}  C[t] > \Theta   \\
      0, & \text{otherwise},
    \end{cases}       
\end{equation}
where $\hat {b}_{T_p}$ is the predicted label, 1 indicates there is a blockage, and 0 means no blockage, the details about how to choose the parameters is described in \sref{sec:eva}.

\subsection{Blockage Time Instance Prediction (\textbf{Problem 2})} \label{subsec:non-ML_p2}
In \textbf{Problem 2}, given the processed LiDAR point cloud before the blockage, we estimate the exact time when the incoming object will block the link. Assuming that the incoming blockage approaches the link at a constant speed, the proposed baseline method first uses DBSCAN clustering algorithm \cite{ester1996density} to extract the points that represent the trace of the incoming object, then uses Least Square (LS) algorithm to estimate the speed of the object and finally estimates the location of the object by averaging the position of the object points. Based on the speed and position estimates, the method is able to estimate the time. The details of the method are as follows:

\begin{figure*}
 \centering
 \includegraphics[width=\linewidth]{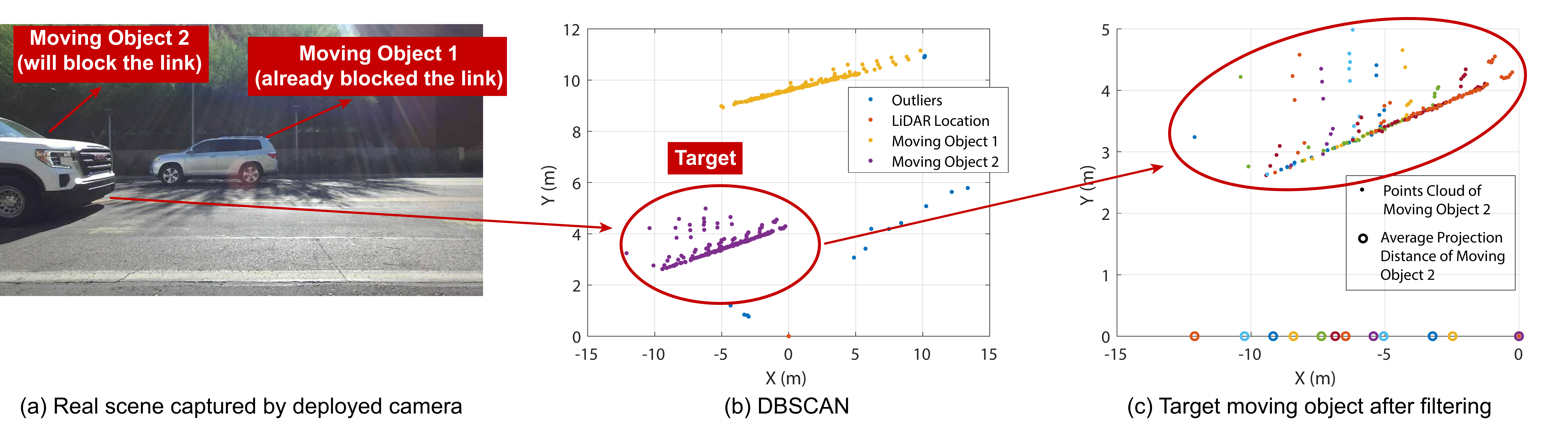}
 \caption{Example of multiple objects in an observation window. (a) shows the real scene captured by the deployed camera where two objects are in the same observation window, (b) shows the LiDAR points classified by DBSCAN, (c) illustrates the LiDAR points of the target (incoming blockage) moving object and its average projection distance on the x-axis, we use different colors for the points that were collected at different time instance. }
 \label{fig:lidar_p2_2object}
\end{figure*}

\subsubsection{DBSCAN} \label{subsubsec:dbscan}
In most cases, there is only one moving object in the point cloud, however, there are some cases where multiple objects occur in the same observation window. Multiple-object case dramatically degrades the performance of estimation since the speed and location information are not accurate. So we use DBSCAN algorithm to cluster the points into different groups and then use a filter to extract the incoming blockage, which we define as \textbf{target object}. Such an example is shown in \figref{fig:lidar_p2_2object} and the real scene captured by the camera is shown in \figref{fig:lidar_p2_2object}a. In this example, two  objects are moving in opposite directions. Object 1 has blocked the link already, and object 2 is the incoming blockage, so our target object is object 2. By applying DBSCAN algorithm, we first remove the outliers and cluster the objects; the parameters of DBSCAN are shown in \sref{sec:eva}. After clustering, the next step is to find the target object and extract its trace. 

The moving direction and location of the object are the main pieces of information that are used to distinguish whether the object has already blocked the LoS link or is going to block it. We first determine the moving direction of the object. Then based on the latest position of the object, we determine our target object. In the example shown in \figref{fig:lidar_p2_2object}, the latest position of object 1 is in the left plane and the moving direction is from right to left, \footnote{left plane is the area $x < 0$, and the right plane is the area where $x > 0$} thus we can determine that it has blocked the link already. As for object 2, the latest position is in the left plane and the moving direction is from left to right, and so we expect it to block the link and keep its points while eliminating all other points, as shown in \figref{fig:lidar_p2_2object}c.

\subsubsection{Least Square Estimation} \label{subsubsec:LS}
The objective of this step is to estimate the speed and the initial location of the object with respect to LiDAR location. This helps in estimating the exact time the object blocks the LoS link. We average the position of the points of the target object at every time instance and project them onto the x-axis, shown as circles in \figref{fig:lidar_p2_2object}. Since the origin is LiDAR sensor, and the LoS link is along the y-axis, the averaged projected distance on the x-axis is the distance between the object and LoS link. Next we calculate this distance which is denoted by $\mathbf{Y} = [y[1] ... y[T_{ob}]]$. The speed of the object is represented by $v$ and the initial distance between the object and LOS link is denoted by $b$. The speed model is presented as:
\begin{equation}
    \begin{bmatrix}
\mathbf{t} & \mathbf{I} 
\end{bmatrix}
\begin{bmatrix}
v\\ 
b
\end{bmatrix}
= \mathbf{Y},
\end{equation}
where, $\mathbf{t} = \left[1, ... T_{ob}\right]^T$ is the time sequence vector, $\mathbf{I}$ is the identity matrix. The estimated speed $\hat{v}$ and initial distance between LoS link and object $\hat{b}$ can be calculated as:

\begin{equation}
\begin{bmatrix}
\hat{v}\\ 
\hat{b}
\end{bmatrix}
=
\left (\begin{bmatrix}
\mathbf{t} & \mathbf{I} 
\end{bmatrix}^\intercal
\begin{bmatrix}
\mathbf{t} & \mathbf{I} 
\end{bmatrix} \right )^{-1}
\begin{bmatrix}
\mathbf{t} & \mathbf{I} 
\end{bmatrix}^\intercal
\mathbf{Y},
\end{equation}

The parameters $\hat{v}$, $\hat{b}$, and the latest distance between the object and LoS link $y[T_{ob}]$ are used to estimate the time when the object will block the link.

\subsection{Blockage Severity Level Prediction (\textbf{Problem 3})} \label{subsec:non-ML_p3}
The threshold-based method is proposed to predict the severity level of an object. Since the size of the object in higher severity levels is larger, so in the LiDAR point cloud, there are more points to represent it. By setting a certain number of points as a threshold, the severity level of objects can be predicted. We count the total number of LiDAR points in the observation sequence and compare the number with a selected threshold to classify the severity level. Recall that $C[t]$ is the total number of LiDAR cloud points at time instance $t$. $\Theta_1$, $\Theta_2$, $\dots$, $\Theta_{N}$ are the pre-defined threshold values to determine the severity level of the blockage and $\hat{b}_{\text{sev},v}$ denotes the predicted blockage severity index. The threshold-based method is summarized below:
\begin{equation} \label{equ:p3_non_ML}
	\hat{b}_{\text{sev},v} = \left\{ \begin{array}{ll}
		1,   \sum_{t=1}^{T_{ob}}  C[t] < \Theta_1,\\
		2,      \Theta_1 < \sum_{t=1}^{T_{ob}}C[t] < \Theta_2, \\
		\vdots & \\
		N_\mathrm{class},    \Theta_{N-1} < \sum_{t=1}^{T_{ob}}C[t] < \Theta_{N}. \\
	\end{array} \right.
\end{equation}

\subsection{Blockage Direction Prediction (\textbf{Problem 4})} \label{subsec:non-ML_p4}
The proposed baseline method to decide the moving direction of the object (\textbf{Problem 4}) is based on the location information given by the LiDAR points corresponding to the earliest and latest time instances in the observation window. For example, in \figref{fig:lidar_p2_2object}c, we calculate the average location of a moving object and project it onto the x-axis; in this case, we focus on the direction along the x-axis. In our LiDAR point cloud, the negative value on the x-axis represents the point on the left side of the LiDAR and the positive value for the right side. We determine the moving direction of the object by comparing its earliest and latest location. Let $x_m[t]$ denote the average projection location at time instance $t$, then the motion direction is estimated by:
\begin{equation} \label{equ:p4_non_ML}
    \hat {b}_{\text{dir}} = 
    \begin{cases}
      0, &   x_m[T_{ob}] >  x_m[1]  \\
      1, & \text{otherwise}.
    \end{cases}       
\end{equation}
Recall that $T_{ob}$ is the last time instance in the observation window and $\hat{b}_{\text{dir}}$ is the predicted motion direction of a blocking object.

\section{ML based Methods} 
\label{sec:ML_method}
Given the complex nature of these pre-blockage signatures, we leverage ML (and in particular deep learning models) to utilize them as in \cite{alkhateeb2018machine,alrabeiah2020deep,charan2021vision,DemirhanICC,mmwave_journal,wu2021lidar}. While the proposed baseline LiDAR methods work well for simple scenarios, complex scenarios such as blockage severity level prediction benefit from more advanced ML-based solutions. The following three subsections present our deep neural network-based approach starting from the network choice, architecture, and processing pipeline.

\begin{figure}[t]
	\centering
	\includegraphics[width=1\linewidth]{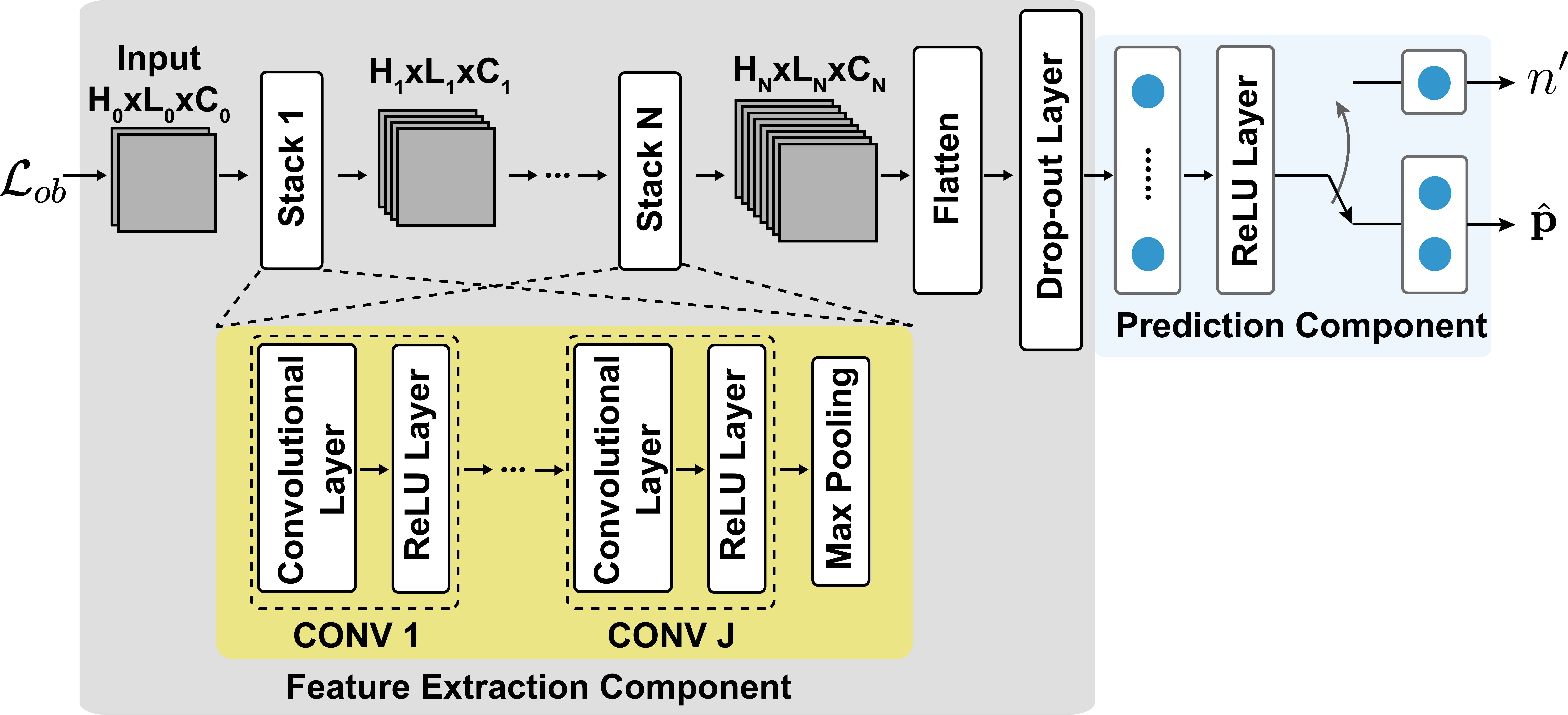}
    \caption{The overall CNN architecture to predict the link status as well as the type and moving direction of the blockage. The two main components are (i) the feature extraction component, and (ii) the prediction component.}
    \label{fig:lidar_cnn_arch}
\end{figure}

\begin{figure*}[t]
	\centering
	\includegraphics[width=1\linewidth]{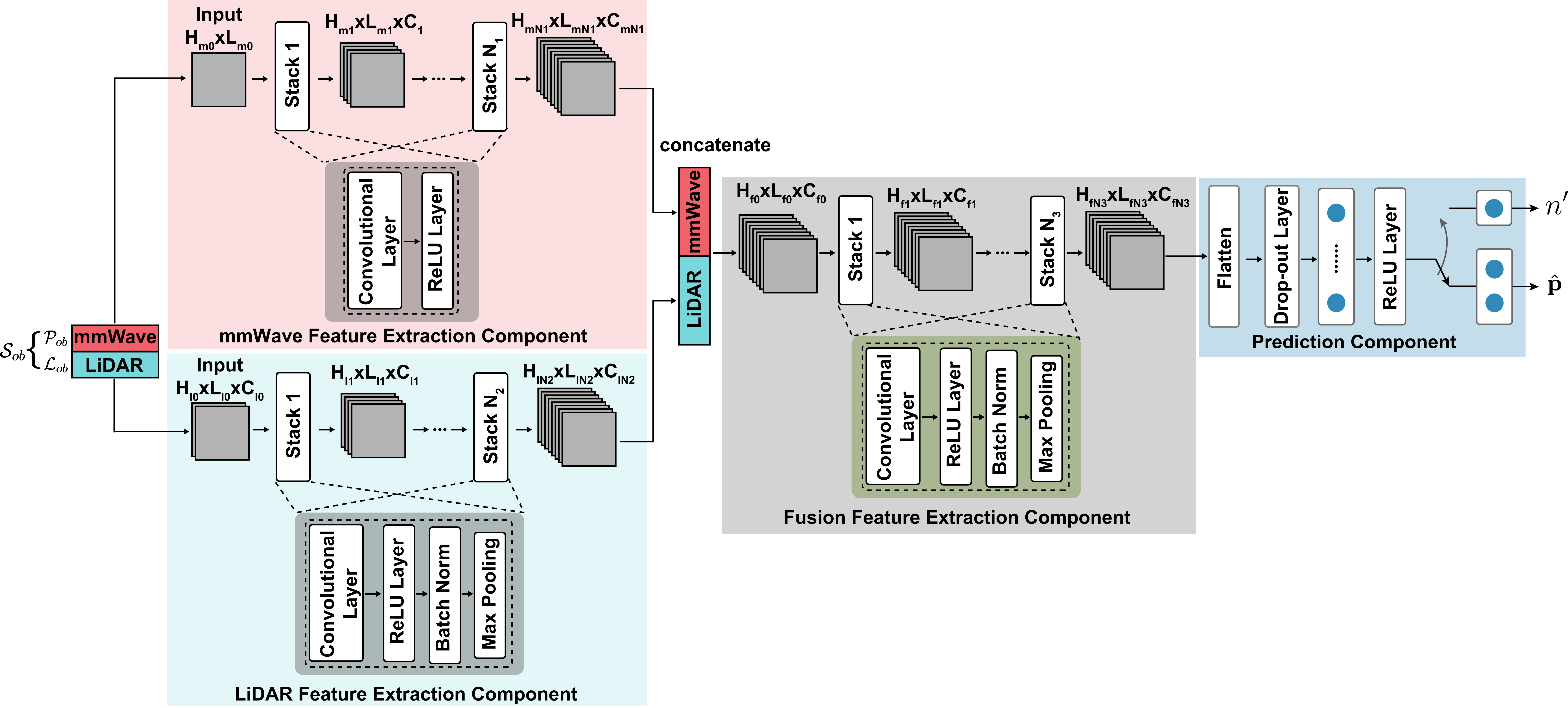}
    \caption{The overall multi-modal architecture to predict the link status as well as severity level and moving direction of the blockage, four main components are shown in the architecture:(i) the mmWave feature extraction component, (ii) the LiDAR feature extraction component, (iii) the fusion feature extraction component, and (iv) the prediction component.}
    \label{fig:multi_arch}
\end{figure*}

\subsection{Convolutional Neural Network Architecture}
\label{subsec:CNN}
To learn the pre-blockage signature at the basestation, we designed the CNN model in \cite{resnet,Yolo} depicted in \figref{fig:lidar_cnn_arch}. Here, $H$, $L$ and $C$ denote the number of rows, columns, and channels of the input data. The CNN architecture consists of a feature extraction component followed by a prediction component. In the feature extraction component, the first stack (Stack 1) takes the input collected LiDAR data (i.e., $\mathcal L_{ob}$) whose dimension is $H_0 \times L_0 \times C_0$ and passes the output with the dimension $H_1 \times L_1 \times C_1$ to the next stack. We design $N$ similar stacks and the last stack (Stack $N$) is followed by a flatten layer (a flatten layer converts the 2-D matrix to 1-D vector). Each stack contains $J$ convolutional blocks, each of which consists of a convolutional layer and a ReLU layer, and a max-pooling layer that occurs at the end of the stack. The output of the feature extraction component is fed to a fully connected (FC) layer followed by a classifier. The classifier outputs a probability vector ($\hat{\mathbf p}$) of whether the link status is blocked or not in $T_P$ future time instances.

\subsection{Multi-Modal Neural Network Architecture}
\label{subsec:multi-modal}
In our earlier work \cite{mmwave_journal}, we showed that mmWave data can predict the blockage with high accuracy when the prediction time interval is short but cannot predict well when the blockage happens further in the future. Further, in \cite{wu2021lidar} we showed that LiDAR has good accuracy for long prediction intervals. So in this paper, we develop a multi-modal neural network for processing fused data consisting of both mmWave wireless and LiDAR data to achieve better performance of blockage prediction for a larger range of prediction intervals. 

The architecture of the proposed multi-modal neural network is shown in \figref{fig:multi_arch}. In the first stage, we input the collected wireless data and LiDAR data to the corresponding feature extraction component, separately. The two components are shown in red and green shadowed areas in \figref{fig:multi_arch}. The mmWave feature extraction component first takes the input mmWave observation sequences whose dimensions are $H_{m0} \times L_{m0}$ and passes the output with dimension $H_{m1} \times L_{m1} \times C_{m1}$ to the next stack. We design $N_1$ similar stacks with the same architecture, each consisting of a convolutional layer and a ReLU layer. 

The LiDAR feature extraction component is similar to the mmWave feature extraction component, except that the LiDAR data is 3-dimensional. The input is $H_{l0} \times L_{l0} \times C_{l0}$. In the LiDAR feature extraction component, each stack consists of a convolutional layer, a ReLU layer, a batch normalization layer, and a max pooling layer. Since the input size of LiDAR is much larger than wireless data, to accelerate the neural network, we use batch normalization \cite{ioffe2015batch}. A max pooling layer is used to adjust the first two dimensions of LiDAR data to be identical to these of wireless data ($H_{l_{N2}} = H_{m_{N1}}$ and $L_{l_{N2}} = L_{m_{N1}}$) in order to concatenate with wireless data in fusion feature extraction component.

The output of the mmWave feature extraction component and LiDAR feature extraction component are concatenated along the third dimension (which represents the number of channels) and input to the fusion feature extraction component. The input dimension is $H_{f0} \times L_{f0} \times C_{f0}$, where $H_{f0} = H_{l_{N2}} = H_{m_{N1}}$, $L_{f0} = L_{l_{N2}} = L_{m_{N1}}$ and $C_{F0} = C_{l_{N2}} + C_{m_{N1}}$. The architecture of the stack in the fusion feature extraction component is identical to that in the LiDAR feature extraction component. It consists of a convolutional layer, a ReLU layer, a batch normalization layer, and a max pooling layer. 

The output of the fusion feature extraction component, whose dimension is $H_{f_{N3}} \times L_{f_{N3}} \times C_{f_{N3}}$, is fed to the prediction component which is identical to the one in \figref{fig:lidar_cnn_arch}. The prediction component contains a flatten layer followed by a drop-out layer, a FC, and a ReLU layer. The output of the ReLU layer is fed to either a classifier or a regressor depending on the problem.

\begin{table}[t]
	\caption{DeepSense Scenario 24 - 27 (site 0)}
	\centering
	\setlength{\tabcolsep}{5pt}
	\renewcommand{\arraystretch}{1.4}
	\begin{tabular}{|c|c|}
		\hline\hline
		\textbf{Testbed}             & 3               \\ \hline
		\textbf{Number of Instances} & \thead{Scenario 24: 40000 - Scenario 25: 80000 \\ Scenario 26: 80000 -Scenario 27: 100000\\ Combined:  300000 (from 3436 trajectories)   }                    \\ \hline
		\textbf{Number of Units}     & 2 \\ \hline
		\textbf{Data Modalities}     & \thead{RGB images, LiDAR point cloud, positions, \\ mmWave beam training measurements} \\ \hline \hline
		\multicolumn{2}{|c|}{\textbf{Unit 1: Stationary}} \\ \hline
		\textbf{Hardware elements} & \thead{RGB Camera, \\  mmWave receiver with 16-element \\ phased array, LiDAR} \\ \hline
		\textbf{Data Modalities} & \thead{RGB images, \\ LiDAR point cloud, GPS position, \\ mmWave beam training measurements}  \\
		\hline \hline
		\multicolumn{2}{|c|}{\textbf{Unit 2: Stationary}} \\ \hline
		\textbf{Hardware elements} &  \thead{mmWave transmitter with an omni-antenna }  \\ \hline
		\textbf{Data Modalities} & GPS position \\
		\hline\hline
	\end{tabular}
	\label{tbl:site0}
\end{table}

\subsection{Training Loss}
For the {{prediction problems (Problems 1, 3, 4)}}, predicting the future link status, blockage type, and moving direction are posed as classification problems, in which the classifier attempts to determine the link status, the type of blockages or the moving direction of the blockages for a future time interval. As such, the network training is performed with a cross-entropy loss function $l_{\text{CH}}$ computed over the outputs of the network \cite{DLBook} 
\begin{equation}
	l_{\text{CH}} =  \sum_{k = 1}^{K} p_{k}\log{\hat{p}_{k}},
\end{equation}
where $K$ is the total number of categories. For each problem, $\mathbf p = [p_1, p_2, ...,p_K]^T$ is a one-hot vector, $p_{k}$ is a binary variable corresponding to the $k$th category. The category with the highest probability is encoded as 1 others are encoded as 0's. For example, if there are three severity levels, then ${b}_{\text{sev},v} = 1, 2, 3$ corresponding to severity levels of 1, 2, and 3. As an example ${b}_{\text{sev},v} = 1$ is represented by $[1,0,0]^T$.

For the {{regression problem (Problem 2)}}, we pose it as a problem of predicting the blockage instance. Our model tries to determine the exact time instance at which the blockage occurs. We use Mean Square Error (MSE) loss as training function. In formal terms, we aim to minimize the difference between the predicted instance and ground truth instance \cite{DLBook}:
\begin{equation}
	l_{\text{MSE}} =  (n^{\prime (u)}- \hat{n}^{\prime (u)})^2,
\end{equation}
where $n^{\prime (u)}$ and $\hat{n}^{\prime (u)}$ are ground truth time instance and predicted time instance, respectively.

\begin{table}[t]
	\caption{DeepSense Scenario 28 (site 1)}
	\centering
	\setlength{\tabcolsep}{5pt}
	\renewcommand{\arraystretch}{1.4}
	\begin{tabular}{|c|c|}
		\hline\hline
		\textbf{Testbed}             & 3               \\ \hline
		\textbf{Number of Instances} & \thead{Scenario 28: 100000 (from 910 trajectories) }                    \\ \hline
		\textbf{Number of Units}     & 2 \\ \hline
		\textbf{Data Modalities}     & \thead{RGB images, LiDAR point cloud, positions, \\ mmWave beam training measurements} \\ \hline \hline
		\multicolumn{2}{|c|}{\textbf{Unit 1: Stationary}} \\ \hline
		\textbf{Hardware elements} & \thead{RGB Camera, \\  mmWave receiver with 16-element \\ phased array, LiDAR} \\ \hline
		\textbf{Data Modalities} & \thead{RGB images, \\ LiDAR point cloud, GPS position, \\ mmWave beam training measurements}  \\
		\hline \hline
		\multicolumn{2}{|c|}{\textbf{Unit 2: Stationary}} \\ \hline
		\textbf{Hardware elements} &  \thead{mmWave transmitter with an omni-antenna }  \\ \hline
		\textbf{Data Modalities} & GPS position \\
		\hline\hline
	\end{tabular}
	\label{tbl:site1}
\end{table}

\begin{table}[t]
	\caption{DeepSense Scenario 29 (site 2)}
	\centering
	\setlength{\tabcolsep}{5pt}
	\renewcommand{\arraystretch}{1.4}
	\begin{tabular}{|c|c|}
		\hline\hline
		\textbf{Testbed}             & 3               \\ \hline
		\textbf{Number of Instances} & \thead{Scenario 29: 100000 (from 1022 trajectories) }                    \\ \hline
		\textbf{Number of Units}     & 2 \\ \hline
		\textbf{Data Modalities}     & \thead{RGB images, LiDAR point cloud, positions, \\ mmWave beam training measurements} \\ \hline \hline
		\multicolumn{2}{|c|}{\textbf{Unit 1: Stationary}} \\ \hline
		\textbf{Hardware elements} & \thead{RGB Camera, \\  mmWave receiver with 16-element \\ phased array, LiDAR} \\ \hline
		\textbf{Data Modalities} & \thead{RGB images, \\ LiDAR point cloud, GPS position, \\ mmWave beam training measurements}  \\
		\hline \hline
		\multicolumn{2}{|c|}{\textbf{Unit 2: Stationary}} \\ \hline
		\textbf{Hardware elements} &  \thead{mmWave transmitter with an omni-antenna }  \\ \hline
		\textbf{Data Modalities} & GPS position \\
		\hline\hline
	\end{tabular}
	\label{tbl:site2}
\end{table}

\section{Experimental Setup and Scenarios} \label{sec:exp_testbed}
To evaluate the performance of the proposed approach in real-world environments, we generate measurement-based datasets, following the footsteps of the DeepSense 6G dataset \cite{DeepSense}. We deploy a testbed in an outdoor wireless environment to collect real-world multimodal measurements and construct, what will be henceforth called, \textit{seed datasets}. 
	
\subsection{Testbed Description} \label{subsec:testbed}
The DeepSense 6G dataset framework \cite{DeepSense} defines a generic structure for sensing/communication datasets where a number of units, each equipped with a set of sensors, collect co-existing sensory/communication data. We adopt the \textbf{DeepSense Testbed 3} \cite{mmwave_journal} and add a synchronized LiDAR sensor. \figref{fig:exp_setup} shows the \textbf{DeepSense Testbed 3} which consists of two stationary units, namely Unit 1 and Unit 2. Unit 1 collects mmWave beam training measurements, visual data, and LiDAR data while Unit 2 is equipped with a mmWave transmitter. For each data sample, the system collects a number of measurements including a LiDAR sample and an RGB image. The scanning range of the LiDAR is 16 meters and the motor spin frequency is 10Hz.

\subsection{DeepSense Scenarios 24-29} \label{subsec:scenario}
\begin{figure*}[t]
	\centering
	\includegraphics[width=1\linewidth]{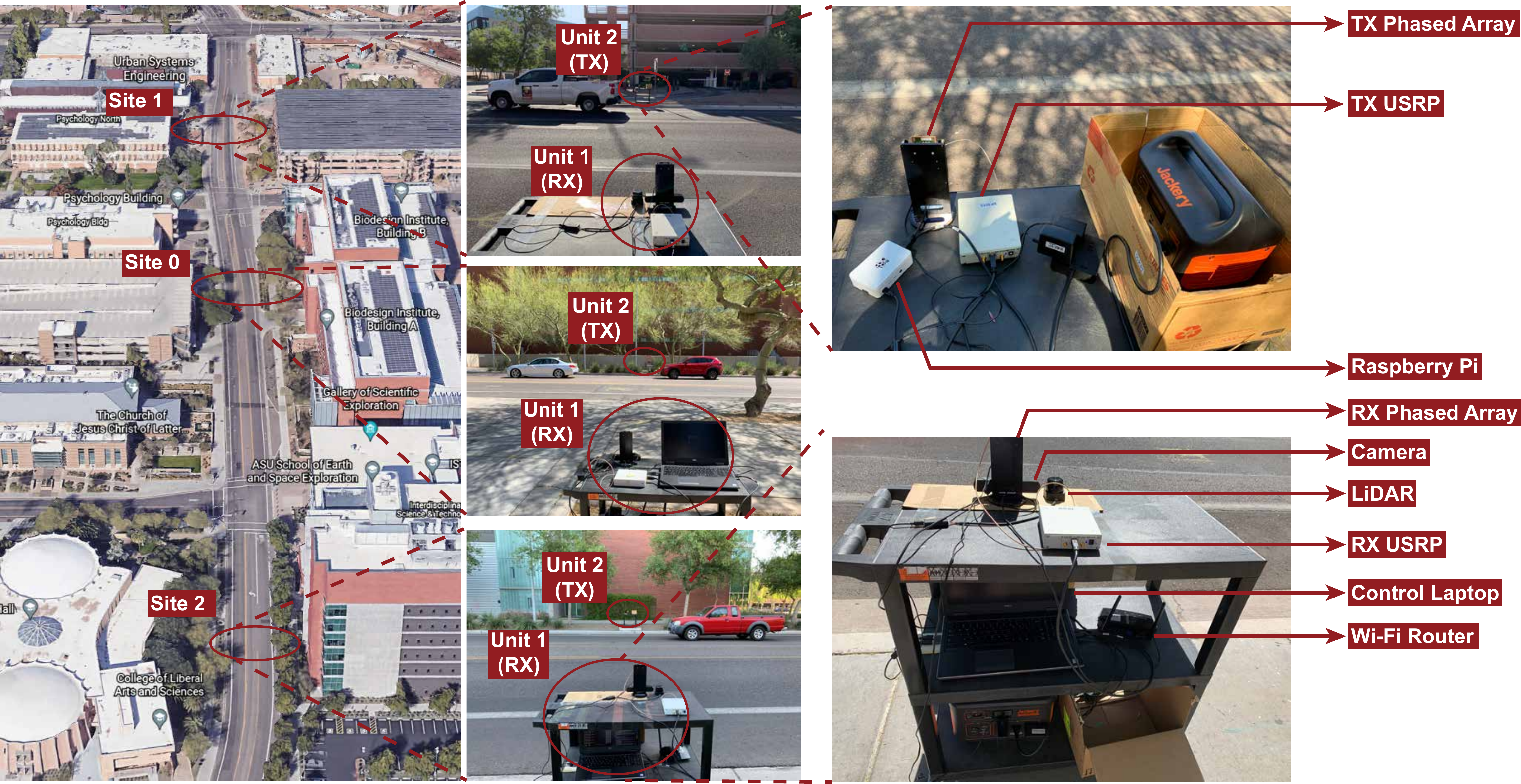}
	
    \caption{The data collection setup of Scenarios 24-29. The left subfigure shows the exact three locations where the data is collected; the middle subfigure shows the traffic and street view from the Unit 1 perspective; the upper plot in the right subfigure shows the hardware setup at Unit 2, the lower figure in the right subfigure shows the hardware setup at Unit 1.}
    \label{fig:exp_setup}
\end{figure*}

We collect data in an outdoor wireless environment representing a two-way city street, as shown in \figref{fig:exp_setup}. The two units of \textbf{DeepSense Testbed 3} are placed on two sides of the street. The LiDAR at Unit 1 continuously scans the environment. The testbed collects data samples at a rate of 10 samples/s. Each data sample has multiple modalities including an RGB image and a LiDAR 360-degree point cloud, both collected by Unit 1. We collect the majority of data in site 0 (300,000 samples) and 100,000 samples in site 1 and site 2, respectively. The important aspects of these DeepSense scenarios are summarized in \tabref{tbl:site0}, \tabref{tbl:site1}, and \tabref{tbl:site2}.

\section{Pre-processing and Development Dataset Generation} \label{sec:dev_dataset}

\subsection{Pre-processing for wireless data}
\label{subsec:pre-wireless}
For wireless data pre-processing, we follow the steps presented in our earlier work in \cite{mmwave_journal, DeepSense} and summarize here for completeness. We standardize the inputs by subtracting the mean of the dataset and dividing it by its standard deviation. 

\subsection{Pre-processing for LiDAR Data (SCR)}
\label{subsec:pre-lidar}

\begin{figure*}[tb]
	\centering
	\begin{subfigure}[t]{0.25\textwidth}
		\centering
		\includegraphics[width=\linewidth]{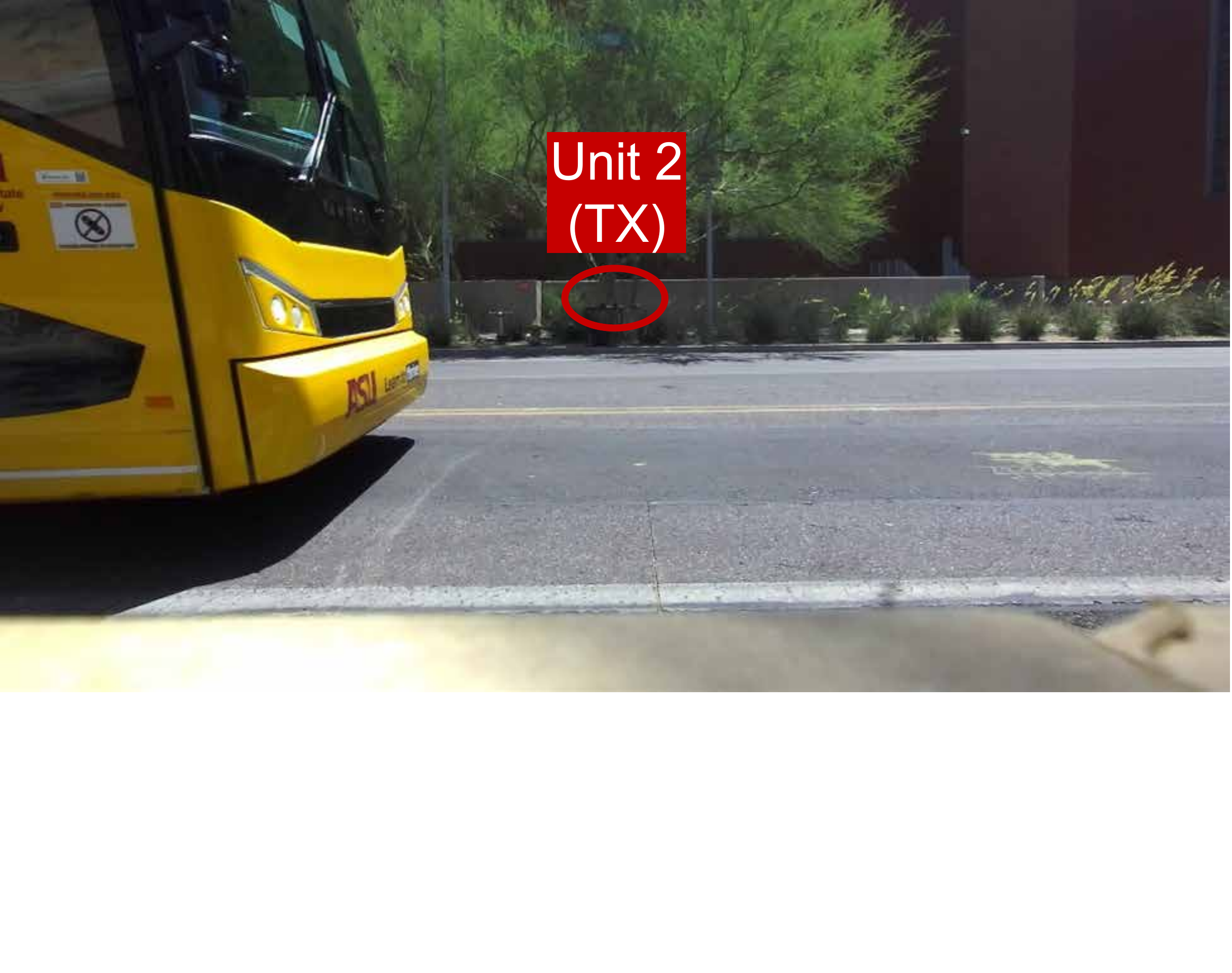}
		\caption{}
		\label{fig:SCR_object}
	\end{subfigure}
	\begin{subfigure}[t]{0.35\textwidth}
		\centering
		\includegraphics[width=\linewidth]{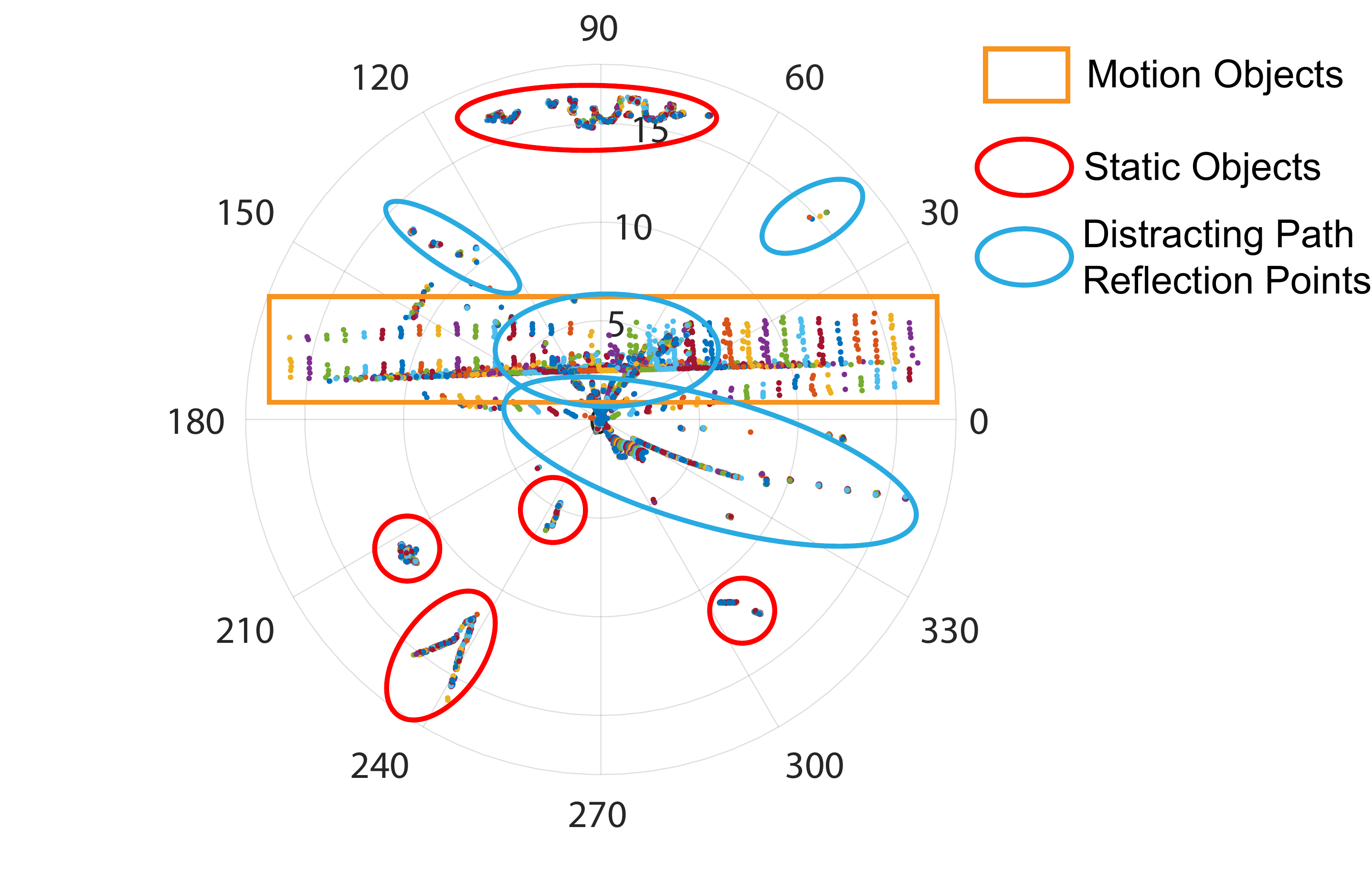}
		\caption{}
		\label{fig:SCR_ori}
	\end{subfigure}
	\begin{subfigure}[t]{0.35\textwidth}
		\centering
		\includegraphics[width=\linewidth]{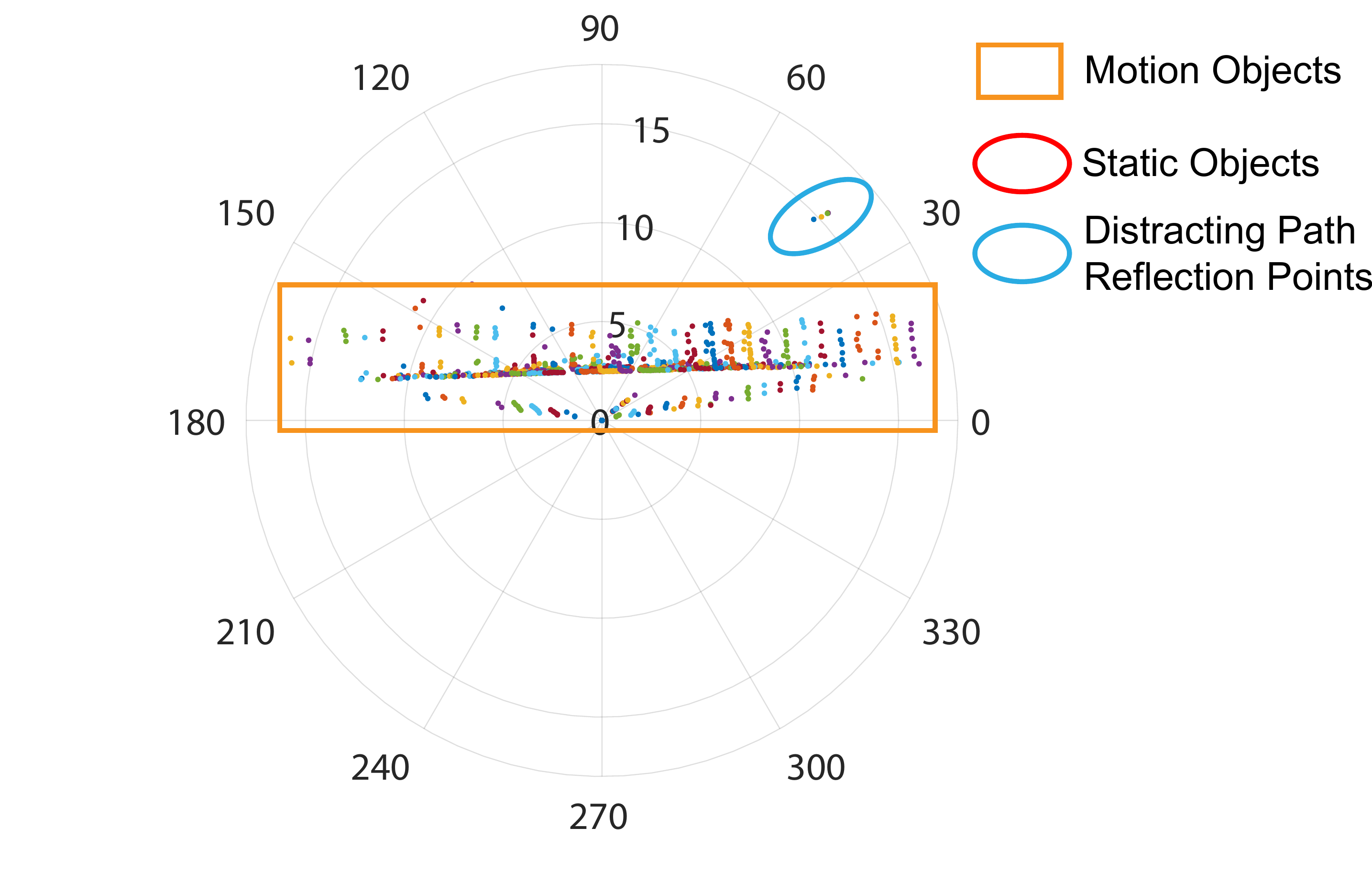}
		\caption{}
		\label{fig:SCR_final}
	\end{subfigure}
	\caption{(a) A moving blockage captured by the RGB camera. (b) The point cloud generated using the raw LiDAR data. It contains the  trace of the moving object, static objects, and distracting path reflection points. (c) The LiDAR  point cloud after  SCR processing. It contains the trace of the moving object and fewer distracting path reflection points.}
	\label{fig:SCR}
\end{figure*}

In order to build the development dataset used in the LiDAR-aided blockage prediction ML task, the raw LiDAR data described in \sref{subsec:scenario} needs to be first pre-processed to (i) remove the noise created by static clusters and (ii) remove the sensory data collected from outside the communication field of view. The details are presented in \cite{wu2021lidar}. 

\figref{fig:SCR_object} and \figref{fig:SCR_ori} show the moving object captured by the RGB camera and the corresponding point clouds of the raw LiDAR data. The points in the orange rectangular represent the trace of moving objects, the ones in the red circles represent the static clusters or objects in the scenario, while the points in the blue circles represent the noisy points that are not present in the real scenario, these are called distracting path reflection points (also static noise). We are interested in the trace of moving objects and so the static objects and the path reflection noise should be removed.

The aim of pre-processing is mainly to eliminate the cluster of static noise points. We first use a field of view-based filtering to erase the LiDAR sensory data collected from directions outside the field of view of interest (\sref{subsec:field_filter}). Next, we use a dictionary-based cluster removal method to remove the unnecessary clusters in \sref{subsec:scr}.

\subsection{Field of View Based Filtering} \label{subsec:field_filter}
Since the objects of interest are between the transmitter and receiver, any LiDAR-detected object on the other side of this communication link needs to be filtered out so that it does not distract the blockage prediction model. Assuming that the LiDAR device collects $P$ samples at every time instance, and each LiDAR sample has 2 values, angle $\phi$ in radians and distance $d$ in meters, the set of samples is $\mathcal L^{(t)} = \{(\phi,d)_p\}_{p=1}^P$, $\mathcal L$ denotes the raw LiDAR dataset, $t$ is the index of time instance. We use the field of view filter to clean the points outside the range ${\Phi}_1$ to ${\Phi}_2$. In this paper, we define ${\Phi}_1 = -\pi/6$, ${\Phi}_2 = \pi$ based on real measurements. We choose $P = 460$ at each time instance since our LiDAR sensor collects 460 samples for a 360-degree point cloud.

\subsection{Dictionary Based Cluster Removal} \label{subsec:scr}
After the field of view-based filtering, some of the static clusters in the LiDAR point cloud are eliminated. We now develop a static cluster removal (SCR) method to remove the rest. The SCR method is implemented in five steps: i) sorting the LiDAR data by their angle; ii) quantizing the angles; iii) quantizing the distance of the LiDAR data; iv) generating a static cluster dictionary from samples that contain no moving objects, and v) eliminating the static clusters according to the constructed dictionary. Sorting and quantization are needed to establish the mapping between points at different time instances.

\textbf{Step 1 Sorting:} Although the number of collected LiDAR samples at each time instance is the same, they are not ordered by either their angles or distances. So the first step is to sort the LiDAR samples by their angles. In the sorting process, we append the zero-distance points at the end of the sorted important points, since these zero-distance points cannot provide effective information.

\textbf{Step 2 Angle Quantization:} After sorting, the angle is in ascending order. We uniformly quantize the angle space to ensure that the points in one time instance can be mapped to those in other time instances within the same quantization step. This establishes a relation between the points at different time instances. We define $Q$ as the total number of angle quantization levels and $q$ as the index of the angle quantization level. The quantization angle is from ${\Phi}_1$ to ${\Phi}_2$ and the step size is denoted as $\Delta \Phi$. If there are multiple points at the same quantization level, we choose the median index of the points whose angle lay in the same quantization level and discard others. In this paper, we choose $Q = 216$.

\textbf{Step 3 Distance Quantization:} 
After angle quantization, the number of important points for each time instance is $Q$. However, due to the measurement error of the device, the measured distance corresponding to an angle may not be exactly the same from one time instance to the next, so the distance values are also quantized. We choose the total distance quantization levels ($Q_d$) to be 500, the step size is 0.034m which provides sufficient accuracy for our 17m LiDAR range.  

\begin{figure}[t]
\centering
\includegraphics[width=\linewidth]{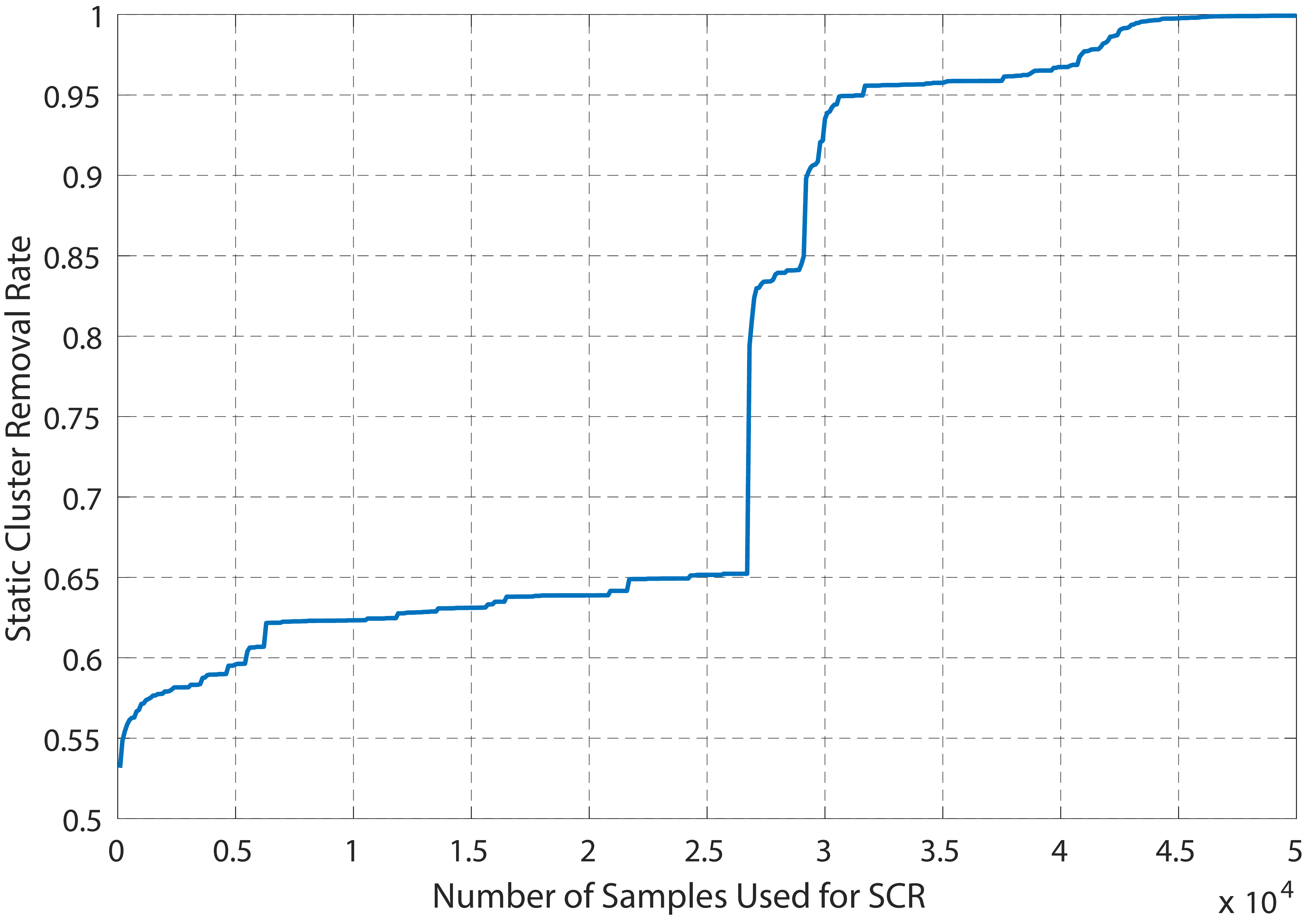}
\caption{Static Cluster Removal (SCR) rate as a function of the number of samples that are used for generating the SCR dictionary.}
\label{fig:scr_perf}
\end{figure}

\textbf{Step 4 Dictionary Generation:} 
For the static cluster dictionary, we choose the samples from $N_d$ time instances ($N_d = 5000$ in our case) which have no moving objects, and remove the repeat points. \figref{fig:scr_perf} illustrates the SCR rate as a function of the number of selected samples to generate the dictionary. The SCR rate is calculated by using the total points that are eliminated by SCR divided by the total points in the original dataset. So the higher rate means the more static cluster is removed. Based on the result of \figref{fig:scr_perf}, the more points to generate the dictionary, the better performance SCR can get. The SCR rate increase sharply from 25,000 to 30,000 samples,  however, in a real-time situation, we can only use the first few samples to generate the dictionary and apply the SCR for the rest of the dataset.

\textbf{Step 5 Residual Cluster Removal:} 
Next, we compare every point in the dataset with every point in the static cluster dictionary. If the point in the dataset is in the static cluster dictionary, it corresponds to a static cluster. Then it gets eliminated by assigning 0 to the distance and we filter out the points with 0 distance. 

\figref{fig:SCR_final} plots the LiDAR point cloud after applying our static cluster removal algorithm. This figure shows that most of the static clusters are removed and the traces of the moving objects are now clear.

\subsection{Development Dataset Generation}
\label{subsec:development_dataset}
 We use the seed dataset of scenario 24-29, described in \sref{subsec:scenario}, to construct the development dataset for the LiDAR-aided blockage prediction task following two steps: (i) Constructing the time sequences from the seed dataset. We follow the same footsteps in \cite{mmwave_journal} to extract the time sequences based on the link status labels. We have 1718 sequences, and each sequence has mmWave receive power, LiDAR data, and the corresponding link status. (ii) Generating development dataset for CNN: we use $\mathcal Y_{P} = \{ (\mathbf {S}_{ob}, b_{T_p})_u \}_{u = 1}^{U}$ to denote the  development dataset. We use \eqref{equ:p1_label} to generate $b_{T_p}$ based on like status, and we apply the sliding window methods \cite{mmwave_journal} to generate $\mathcal S_{ob}$ with time instance length $T_{ob}$. The total number of sequences for the dataset collected in site 0 is $U$ = 3436, 910 in site 1, and 1022 in site 2.

\section{Evaluation Metrics and Parameter Selection} \label{sec:eva}
In this section, we introduce the evaluation metrics used to evaluate the results in \sref{subsec:eva_metric}. \sref{subsec:para_sel} presents the parameters selection for both the baseline method and ML-based method.

\subsection{Evaluation Metrics} 
\label{subsec:eva_metric}
Since {{Problems 1, 3, 4}} are prediction problems, we use Top-1 accuracy as our evaluation metric. It is defined as the compliment of the prediction error given in \cite{ImageNet}, and it is expressed as:
\begin{equation}
	\text{Acc}_{\text{top-1}} = \frac{1}{U_{v1}}\sum_{u = 1}^{U_{v1}} \mathbbm{1}  (b^{(u)} = \hat {b}^{(u)}),
\end{equation}
where $\mathbbm{1}$ is the indicator function, $U_{v1}$ is total samples of the validation set in problem 1, $b^{(u)}$ and $\hat{b}^{(u)}$ are, respectively, the target and predicted link status for a future interval of $T_P$ instances, specifically, $b^{(u)}$ represents  $b_{T_p}^{(u)}$ in {Problem 1}, $b_{\text{type}}$ in {Problem 3} and $b_{\text{dir}}$ in {Problem 4}.

\begin{table*}[]
\centering
\caption{Parameters of CNN for LiDAR data}
\begin{tabular}{cccccccccc}
\cline{1-10}
\hline\hline
\multicolumn{2}{c}{\multirow{3}{*}{\textbf{Name}}} & \multicolumn{8}{c}{ \textbf{Value}}  \\ 
\multicolumn{2}{c}{} &  \multicolumn{4}{c}{\textbf{Original}}    &  \multicolumn{4}{c}{\textbf{SCR}}    \\
\multicolumn{2}{c}{} & \textbf{Prob 1} & \textbf{Prob 2} & \textbf{Prob 3} & \textbf{Prob 4} & \textbf{Prob 1} & \textbf{Prob 2} & \textbf{Prob 3} & \textbf{Prob 4} \\
\hline
\multicolumn{2}{c}{Input Seq. Dim}       &  \multicolumn{4}{c}{16$\times$460$\times$2}  &  \multicolumn{4}{c}{16$\times$216$\times$2}  \\ 
\multicolumn{2}{c}{Predicted future time steps} &  \multicolumn{4}{c}{1-10}  &  \multicolumn{4}{c}{1-10}  \\ 
\multirow{3}{*}{Stack 1}    & Conv 1   &  \multicolumn{4}{c}{2-8-3-1}   &  \multicolumn{4}{c}{2-8-3-1}   \\ 
                            & Conv 2   &  \multicolumn{4}{c}{8-16-3-1}  &  \multicolumn{4}{c}{8-16-3-1}   \\ 
                            &Max pooling 1  &  \multicolumn{4}{c}{(2,23)}    &  \multicolumn{4}{c}{(2,9)}  \\ \hline
\multirow{3}{*}{Stack 2}    & Conv 3   &  \multicolumn{4}{c}{16-16-3-1}  &  \multicolumn{4}{c}{16-16-3-1}  \\ 
                            & Conv 4   &  \multicolumn{4}{c}{16-32-3-1} &  \multicolumn{4}{c}{16-32-3-1}   \\ 
                            & Max pooling 2  &  \multicolumn{4}{c}{(2,5)}    &  \multicolumn{4}{c}{(2,6)}    \\ \hline
\multicolumn{2}{c}{FC}                  &  (512,2)  &  (512,1)  &  (512,3)  &   (512,2)  &  (512,2)  &  (512,1)  &  (512,3)  &  (512,2)  \\ 
\multicolumn{2}{c}{Dropout rate}        &  \multicolumn{4}{c}{0.2}  &  \multicolumn{4}{c}{0.2}    \\ 
\multicolumn{2}{c}{Epoch}               &  \multicolumn{4}{c}{1000} &  \multicolumn{4}{c}{1000}   \\ 
\multicolumn{2}{c}{\multirow{2}{*}{Total Parameters}}    &  \multirow{2}{*}{9306} &  \multirow{2}{*}{8793} & 9306($N_\mathrm{class} = 3$)   &   \multirow{2}{*}{9306}  &  \multirow{2}{*}{6883}    &  \multirow{2}{*}{5557} &    6883($N_\mathrm{class} = 3$)   &  \multirow{2}{*}{6883}\\ 
\multicolumn{2}{c}{}    &   &   & 9819($N_\mathrm{class} = 4$)  &   &     &       & 8209($N_\mathrm{class} = 4$)     &    \\ 

\hline\hline
\end{tabular}
\label{tbl:CNN_lidar}
\end{table*}

\begin{table*}[]
\centering
\caption{Parameters of multi-modal NN for multi-modal data}
\begin{tabular}{cccccccccc}
\cline{1-10}
\hline\hline
\multicolumn{2}{c}{\multirow{3}{*}{\textbf{Name}}} & \multicolumn{8}{c}{ \textbf{Value}}  \\ 
\multicolumn{2}{c}{} &  \multicolumn{4}{c}{\textbf{Original}}    &  \multicolumn{4}{c}{\textbf{SCR}}    \\
\multicolumn{2}{c}{} & \textbf{Prob 1} & \textbf{Prob 2} & \textbf{Prob 3} & \textbf{Prob 4} & \textbf{Prob 1} & \textbf{Prob 2} & \textbf{Prob 3} & \textbf{Prob 4} \\
\hline
\multicolumn{2}{c}{LiDAR Input Seq Dim}       &  \multicolumn{4}{c}{16$\times$460$\times$2}  &  \multicolumn{4}{c}{16$\times$216$\times$2}    \\ 
\multicolumn{2}{c}{mmWave Input Seq Dim}       &  \multicolumn{4}{c}{16$\times$54}   &  \multicolumn{4}{c}{16$\times$54}    \\ 
\multicolumn{2}{c}{Predicted future time steps} &  \multicolumn{4}{c}{1-10} &  \multicolumn{4}{c}{1-10}  \\ \hline
mmWave Stack 1    & Conv 1   &  \multicolumn{4}{c}{2-4-(1,3)-(0,1)}  &  \multicolumn{4}{c}{1-4-(1,3)-(0,1)}     \\  \hline
mmWave Stack 2    & Conv 2   &  \multicolumn{4}{c}{4-8-(1,3)-(0,1)}  &  \multicolumn{4}{c}{4-8-(1,3)-(0,1)}   \\  \hline
\multirow{2}{*}{LiDAR Stack 1}    & Conv 1   &  \multicolumn{4}{c}{2-4-(1,5)-0}  &  \multicolumn{4}{c}{2-4-(1,5)-0}     \\  
                                & Max Pooling 1  & \multicolumn{4}{c}{(1,2)}    & \multicolumn{4}{c}{(1,2)}     \\ \hline
\multirow{2}{*}{LiDAR Stack 2}    & Conv 2   &  \multicolumn{4}{c}{4-4-(1,5)-0}  &  \multicolumn{4}{c}{4-4-(1,5)-0}   \\ 
                                & Max Pooling 2  & \multicolumn{4}{c}{(1,2)}    & \multicolumn{4}{c}{(1,2)}     \\ \hline
\multirow{2}{*}{LiDAR Stack 3}   & Conv 3    &  \multicolumn{4}{c}{4-8-(1,5)-0}  &  \multicolumn{4}{c}{4-8-(1,5)-0}  \\
                                & Max Pooling 3  & \multicolumn{4}{c}{(1,2)}    & \multicolumn{4}{c}{(1,2)}     \\ \hline
\multirow{2}{*}{Fusion Stack 1}   & Conv 1    &  \multicolumn{4}{c}{16-16-3-1}  &  \multicolumn{4}{c}{16-16-3-1}  \\
                                & Max Pooling 1  & \multicolumn{4}{c}{(2,3)}    & \multicolumn{4}{c}{(2,3)}     \\ \hline
\multirow{2}{*}{Fusion Stack 2}   & Conv 2    &  \multicolumn{4}{c}{16-16-3-1}  &  \multicolumn{4}{c}{16-16-3-1}  \\
                                & Max Pooling 2  & \multicolumn{4}{c}{(2,2)}    & \multicolumn{4}{c}{(2,2)}     \\ \hline
\multicolumn{1}{c}{FC}      &            &  (576,2)  &  (576,1)  &  (576,3) &  (576,2)  &  (576,2)  &  (576,1)  &  (576,3) &  (576,2)   \\ 
\multicolumn{1}{c}{Dropout rate} &        &  \multicolumn{4}{c}{0.2}  &  \multicolumn{4}{c}{0.2}   \\ 
\multicolumn{1}{c}{Epoch}    &           &  1000 &  1800  &  1000 &  1800 &  1000 &  1800 &  1000 &  1800 \\ 
\multicolumn{2}{c}{\multirow{2}{*}{Total Parameters}}    &  \multirow{2}{*}{6306} &  \multirow{2}{*}{5729} & 6306($N_\mathrm{class} = 3$)   &   \multirow{2}{*}{6306}  &  \multirow{2}{*}{6134}    &  \multirow{2}{*}{5557} &    6134($N_\mathrm{class} = 3$)   &  \multirow{2}{*}{6134}\\ 
\multicolumn{2}{c}{}    &   &   & 6883($N_\mathrm{class} = 4$)  &   &     &       & 6711($N_\mathrm{class} = 4$)     &    \\ 

\hline\hline
\end{tabular}
\label{tbl:multi-modal}
\end{table*}

For {{Problem 2}}, we use Mean Absolute Error (MAE) defined as the mean absolute error between the ground-truth value and the predicted value, and its standard deviation to evaluate the quality of our model. For each prediction interval $T_P$, we calculate MAE and its standard deviation. While we use MSE to compute training function loss, we use MAE in the evaluation. Since it is easier to see the difference in time instances.

\begin{equation}
	e^{(u)}_{T_P} = \lvert n^{\prime (u)} - \hat{n}^{\prime (u)} \rvert, \quad  \forall u \in \{1,\dots,U_{v2}\} ,    
\end{equation}
\vspace{-0.4cm}
\begin{equation}
	\bar {e}_{T_P} = \frac{1}{U_{v2}}\sum_{u = 1}^{U_{v2}} \lvert n^{\prime (u)} - \hat{n}^{\prime (u)}  \rvert,
\end{equation}

\begin{equation}
	\text{std}_{T_P} = \left({{\frac {1}{U_{v2}}}\sum _{i=1}^{U_{v2}}\left(e^{(u)}_{T_P}-{ \bar {e}_{T_P}}\right)^{2}}\right)^{\frac{1}{2}},
\end{equation}
where, $e^{(u)}_{T_P}$ is the absolute error for $u$th sample, $\bar {e}_{T_P}$ is the MAE, $ \text{std}_{T_P}$ is the standard deviation of absolute error, $U_{v2}$ is the total numbers of samples in the validation set, $n^{\prime (u)}$ and $\hat{n}^{\prime (u)}$ are the target and predicted time instances between the current time and the time of blockage occurrence, assuming prediction interval is $T_P$.

\subsection{Parameter Selection} \label{subsec:para_sel}
In this section, we explain how we chose the parameters for the baseline method and the ML method in different data collection locations.

\subsubsection{Baseline LiDAR-based Method} \label{baseline method}
\textbf{Blockage Occurrence Prediction (Problem 1): }We compute a distinct threshold value for each of the 10 prediction time intervals in site 0. This is done by sweeping the threshold from 1 to the maximum number of points to find the value which can be used to predict the occurrence of blockage with the highest accuracy. So for prediction interval 1-10, the threshold vector is $[14,18,6,3,1,1,1,1,1,1]$. The threshold vector is computed only using data collected in site 0 and is applied to sites 1 and 2.

\noindent\textbf{Blockage Time Instance Prediction (Problem 2): } For DBSCAN \cite{ester1996density}, we chose Epsilon to be 2.1, and the minimum number of neighbors required for the core point to be 10. The parameters are chosen to give the best performance.

\noindent\textbf{Blockage Severity Level Prediction (Problem 3): } For this problem, we follow the same procedure as in \textbf{Problem 1}. We sweep the threshold values and find the one that predicts the severity level with the highest accuracy using the data collected in site 0. The same threshold vector is applied for data collected in sites 1 and 2.

\noindent\textbf{Blockage Direction Prediction (Problem 4): } For this problem, we used the LiDAR points in 1 (first) and 10 (last) time instances to predict the motion direction.

\subsubsection{ML Method} \label{ML method}
The hyper-parameters and parameters of each layer of our CNN model for \textbf{Problems 1, 2, 3 and 4} are shown in \tabref{tbl:CNN_lidar} (for LiDAR data only) and in \tabref{tbl:multi-modal} (for LiDAR and mmWave data). Since after SCR processing, the number of LiDAR data samples per time instance changes, some of the layer parameters change, as shown in \tabref{tbl:CNN_lidar}. We represent the parameter format as input channel - output channel - kernel size - padding, and for max pooling layer, kernel size. To make the size of RNN and CNN models comparable, we use a similar number of weights for each of the networks. All parameters are based on empirical experiments.

\section{Experimental Results} \label{sec:results}
In this section, we present and analyze the simulated results for all four problems using both baseline and ML-based methods in \sref{subsec:result_p1}, \sref{subsec:result_p2}, \sref{subsec:result_p3}, \sref{subsec:result_p4}, respectively. We compare the baseline method with ML-based method at the end of each section.

\subsection{Blockage Occurrence Prediction (\textbf{Problem 1})} \label{subsec:result_p1}

\begin{figure*}[t]
\centering
\includegraphics[width=\linewidth]{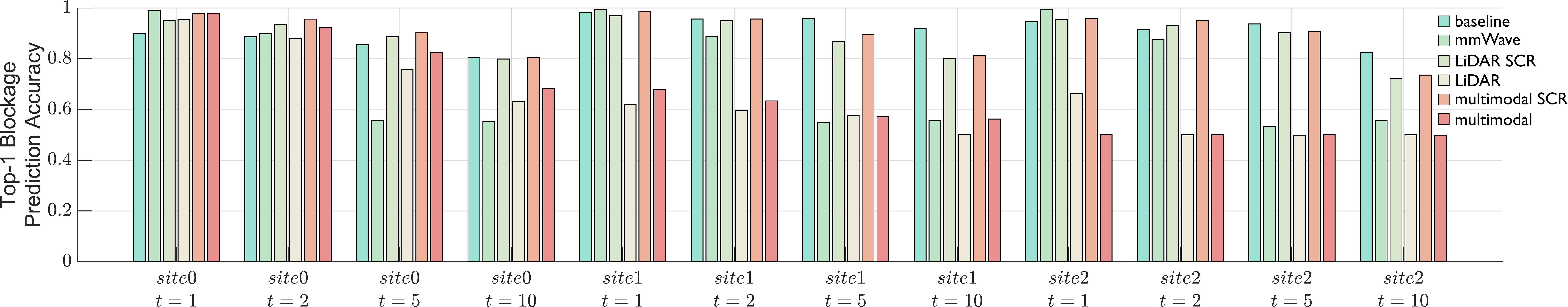}
\caption{Performance of blockage occurrence prediction for prediction time interval 1, 2, 5, 10 time instances (duration of 1 time instance is 100 ms) using proposed baseline LiDAR method and ML-based method by inputting data collected at three locations (site 0, 1 and 2).}
\label{fig:p1}
\end{figure*}

We evaluate the accuracy to predict the occurrence of blockage, for the baseline method and the ML-based method using different data sources (mmWave, LiDAR, LiDAR with SCR, mmWave, and LiDAR, mmWave and LiDAR with SCR). \figref{fig:p1} presents the prediction accuracy results. We choose 4 typical prediction intervals: 1, 2, 5, and 10 time instance, where the duration of 1 time instance is 0.1 s. We show the results using the data collected at all three locations. 

\subsubsection{Baseline Method}
Using the baseline method, the accuracy for all three datasets decrease as the prediction interval (the time before the blockage actually happens) increases. This selected threshold can achieve 90\% when applied in site 0, 98\% in site 1, and 95\% in site 2. And the performance is above 80\% for all three sites when the prediction time interval is 10 time instance (1 s), especially, above 90\% in site 1. This threshold method highly depends on the performance of SCR. Most of the static clusters are eliminated by this algorithm and so a single threshold can be used to predict the incoming blockage.

\subsubsection{ML Based Method}
We first evaluate the performance of our ML-based method by training it and testing it at the same site, namely site 0. We show the potential gain in terms of hand-off latency and finally, we demonstrate the robustness of our model by applying the model to the dataset collected at different locations.

\textit{Without SCR processing:} By comparing the performance of models using wireless and LiDAR data, we see that the model using wireless data outperforms the one using LiDAR data for the first 2 time instances (200 ms) and then the performance drops sharply to 55\% and remains flat. The reason for this is that the length of the wireless signature is short -- 5 time instance (500 ms) in length, whereas the LiDAR signature is much longer (10 time instances or 1 s). The performance of LiDAR data degrades more slowly. Thus with fused input data, our multi-modal model is able to predict the incoming blockage above 90\% accuracy for first 2 time instances (200 ms). The performance of using fused input data is better than the performance of only using LiDAR data with a 5\% better performance on average.

\textit{With SCR processing:} The performance of both multi-modal and LiDAR model improves with SCR processing. Both models can predict the incoming blockage with above 80\% accuracy even for prediction interval of 10 time instances (1 s) before the link is blocked. This guarantees enough time for basestation to hand-off. The performance of the multi-modal system is even better than LiDAR only model with an accuracy improvement of 3\%. Compared to models without SCR processing, the performance improves a lot: for LiDAR only model, the improvement is up to 17\% in accuracy; for multi-modal the accuracy improvement is up to 15\%. 

So we conclude that i) the performance of model using wireless data is best for the first few time instances (1-2); ii) the model input with LiDAR and fused data predicts the blockage with relatively higher accuracy compared to the model using wireless data, and the performance of the model using fused data is better than the model using a single source (wireless or LiDAR). iii) Multi-modal with SCR-processed LiDAR data has the best performance and thus it is a promising solution to solve the blockage prediction problem.

\begin{figure}[t]
\centering
\includegraphics[width=\linewidth]{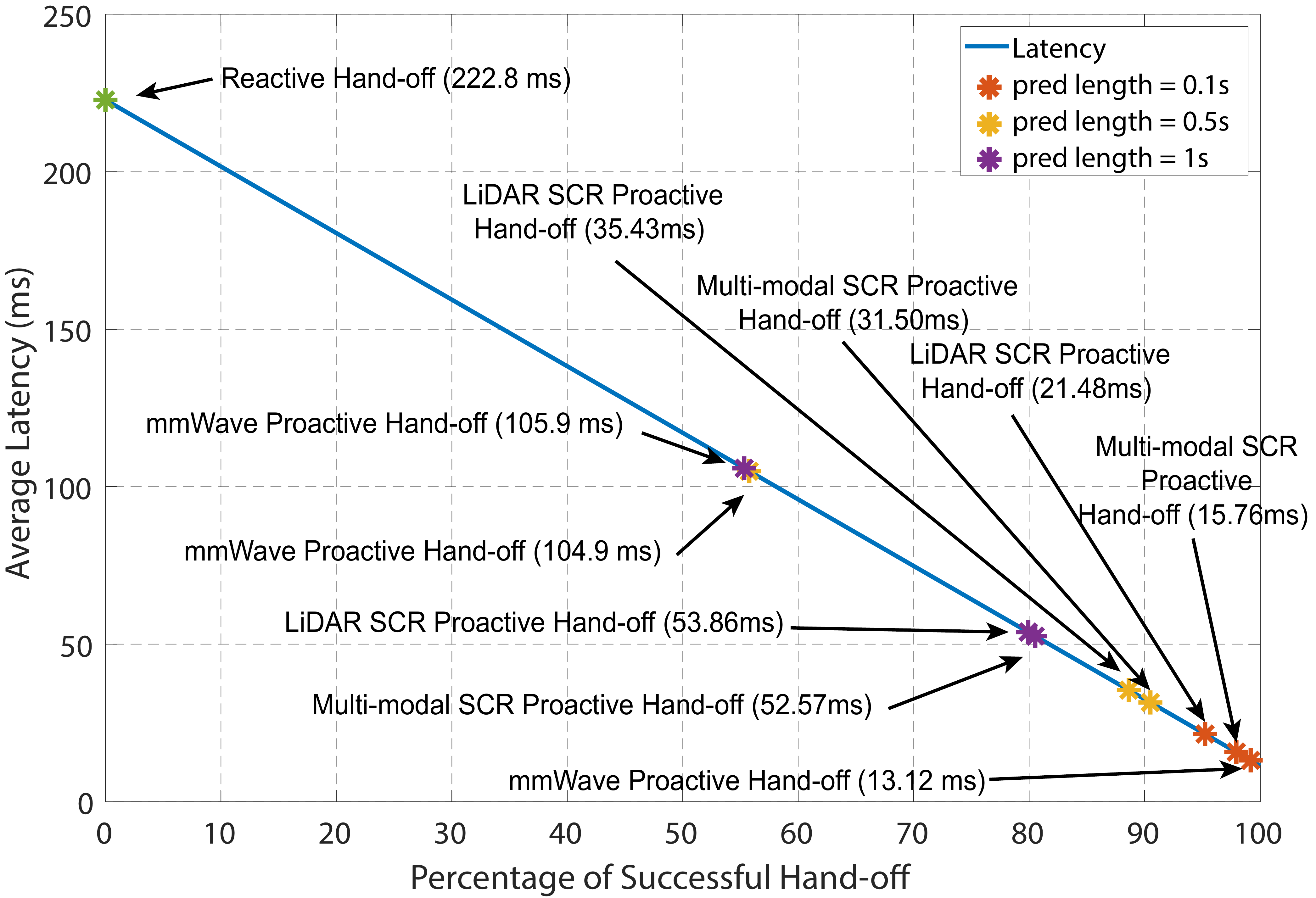}
\caption{Average latency of the proposed proactive hand-off solutions compared to the conventional reactive approaches.}
\label{fig:latency}
\end{figure}

\noindent\textbf{Latency Analysis:} To draw some insights about the potential gains of the proposed methods for the initial access latency in 3GPP 5G NR, we adopt the approach in \cite{charan2021vision} for analyzing the results. According to the 3GPP specifications, a conventional reactive hand-off results in an overall delay of 222.8 ms. If the blockage is predicted proactively, a successful proactive hand-off scenario results in 11.4ms latency associated with the contention free random access \cite{charan2021vision}. Based on the prediction accuracy and the latency given by the average latency for the user hand-off, the average latency $\delta$ for the user is given by $\delta = \hat{p} \times 11.4 + (1-\hat{p}) \times 222.8$,  where, $\hat{p}$ is the blockage prediction accuracy at 100 ms, 500 ms, and 1 s.
	
\figref{fig:latency} shows the average latency improvement of the proposed methods compared to reactive hand-off. By applying our proposed methods, we can achieve an average latency of 13.12 ms using wireless data, 15.76 ms using fused LiDAR and wireless data, and 21.48 ms using LiDAR SCR data when the prediction length is 100 ms. As the prediction interval increases, the LiDAR SCR based approach consistently maintains low latency compared to the other two solutions. Compared to the reactive hand-off latency, our approaches realize more than 10 times improvement in latency.

\noindent\textbf{Robustness Analysis:} To test the robustness of our proposed models, we trained the models in site 0 and then tested our models using the datasets collected in sites 1 and 2; the corresponding results are shown in \figref{fig:p1}, respectively. 

\textit{Without SCR processing}: We see from \figref{fig:p1} that without SCR, the model with LiDAR and multi-modal do not perform well. In site 1 the highest accuracy for multi-modal is 67\%, and for LiDAR model it is 63\%. The performance is worse in site 2, where both models fail to predict the blockage. Since the LiDAR data provides position information, it includes all objects in the environment (both static clusters and moving objects). In addition, different sites have different static cluster patterns, and so even if the model learns to detect the LiDAR signatures among the static clusters in one site, it is not able to predict the blockage when the data changes resulting in different static cluster patterns. 

\textit{With SCR processing}: Most of the resident static clusters in the LiDAR dataset are eliminated with SCR processing. The models trained in site 0 are able to predict the blockage in other sites by recognizing the LiDAR signatures. In \figref{fig:p1}, the performance of models using SCR-processed LiDAR data improve a lot compared to the ones without SCR processing. And the performance in site 1 is comparable to the performance in site 0, which means the model is robust in site 1. The performance in site 2 degrades a little bit compared to sites 0 and 1. The static clusters in site 2 are complex, and so the SCR algorithm is not able to completely clean all the static clusters. 

Unlike the models that use LiDAR data, the performance of the model using wireless data doesn't change much at both sites. Since the wireless signature is based on energy type information formed due to the moving blockage, the model is able to predict the blockage at different sites. This explains the fact that the performance of the model using wireless data doesn't change much when the input data comes  from different sites. However, the performance of the LiDAR model and multi-modal model drops dramatically when the input data comes from different sites.

\begin{table}[]
\centering
\caption{The blockage prediction accuracy when prediction time interval = 5 (500 ms) using ML method.}
\begin{tabular}{cccc}
\hline \hline
\multirow{2}{*}{Data modality} & \multicolumn{3}{c}{Problem 1} \\ 
                               & site 0   & site 1   & site 2  \\ \hline
mmWave                         & 56\%      & 55\%      & 56\%     \\
mmWave + LiDAR                 & 83\%      & 57\%      & 50\%     \\
mmWave + LiDAR SCR             & 90\%      & 90\%      & 90\%   \\ \hline \hline
\end{tabular}

\label{tab:p1}
\end{table}

\noindent\textbf{Summary:} We conclude that the ML model using wireless data performs well for the first two time instances (200 ms), but the performance degrades sharply after that. It is robust in the sense that it has similar performance in different sites. The models using LiDAR data without SCR and fused data perform well at the site where they are trained, but the performance degrades when the model is trained at one site and tested on another unseen scenario. However, with SCR-processed input data, the models perform well in all the other unseen scenarios. It improves accuracy in site 0 by up to 15\%, and makes it possible to predict the blockages in different sites with a performance compatible to that in site 0. \tabref{tab:p1} summarizes the performance of the models using mmWave data, mmWave and LiDAR data with or without SCR in three locations when the prediction time interval is 5 time instances (500 ms). With SCR processing, the performance improves by up to 40\% compared to the performance without SCR processing for unseen scenarios. On the flipside, the SCR algorithm takes much more processing time compared to the collection time (10 times each sample). Thus SCR based methods in their current form cannot be used for real-time prediction.

With SCR processing, the majority of static clusters and irrelevant moving objects are removed, and the LiDAR signatures are cleaner. The problem then becomes simple, all we need is an algorithm to recognize the LiDAR signatures. We find that the threshold-based method is efficient enough to predict the incoming blockage with high accuracy. The ML models are able to achieve high accuracy as well, however, due to their complex architectures, the proposed baseline LiDAR method may be preferred for this problem.

\subsection{Blockage Time Instance Prediction (\textbf{Problem 2})} \label{subsec:result_p2}
\begin{figure*}[t]
\centering
\includegraphics[width=\linewidth]{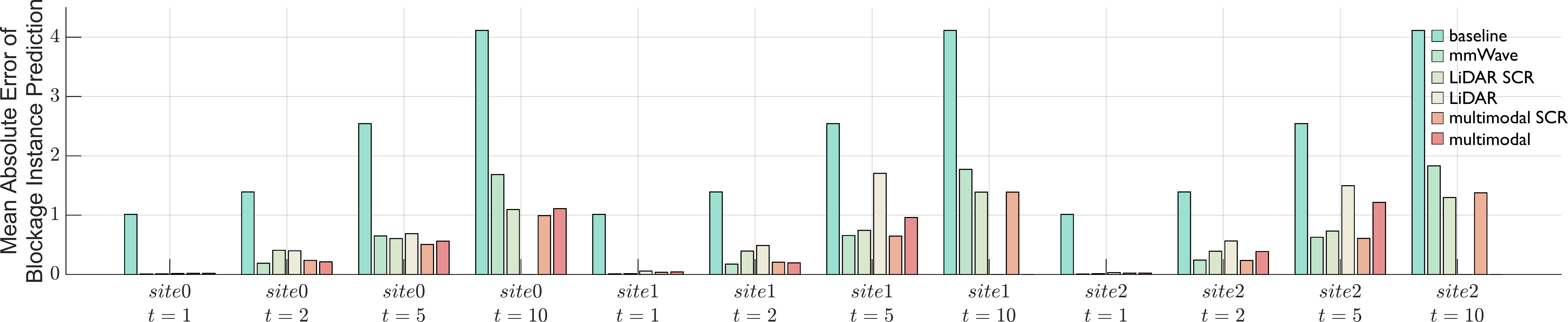}
\caption{Mean absolute error (MAE) between the target (the exact time instance where the blockage happens) and the prediction of this blockage time instance for future prediction intervals 1, 2, 5, 10 time instances (duration of 1 time instance is 100 ms) using proposed baseline LiDAR method and ML based methods for the data collected at all three locations.}
\label{fig:p2}
\end{figure*}

We evaluate the performance of all models for 4 typical prediction intervals, namely, 1, 2, 5, and 10 time instances, in sites 0, 1, and 2. \figref{fig:p2} shows the exact time when LoS link will be blocked using different methods, baseline method, ML-based method that use different data sources (mmWave, LiDAR, LiDAR SCR, mmWave and LiDAR, mmWave and LiDAR SCR).
\noindent\textbf{Baseline Method:} The performance at all three sites of the baseline method is similar; the difference of mean absolute error of blockage at each site is less than 0.2 time instances (20 ms) for the typical 4 prediction time interval (400 ms). The absolute error of estimation is around 1 time instance (100 ms) at first time instance and then as the prediction interval increases, the absolute error increases. The estimation algorithm predicts the exact time when the object blocks the link with 4.1 time instances (410 ms) mean absolute error in site 0, 4.05 time instances (405 ms) in site 1, 3.7 time instances (370 ms) in site 2. The errors are caused by the sampling rate of LiDAR device. At every sample, the number of points that represents the moving blockage is not the same. For example, at time instance 1, 15 points represent the object, but at time instance 2, 8 points represent the object. The variation in the number of collected points representing an object is the main cause of the error.

\noindent\textbf{ML Based Method:} Similar to \textbf{Problem 1}, we trained our models in site 0 and tested them in sites 0, 1, and 2. Due to the limitation of GPUs' memory size, some of the results could not be analyzed for large future prediction intervals. In \figref{fig:p2}, we present the results when the models are trained and tested in site 0. The multi-modal model and model using wireless data have similar performance for the first 2 time instances (200 ms). As the number of time instances increases, the absolute error of the model using wireless data increases while the performance of multi-modals models are still good. Since multi-modal models take both LiDAR data and wireless data, it takes advantage of both data sources. Thus multi-modal model outperforms the models using LiDAR data for the first 2 time instances (200 ms) with and without SCR pre-processing.

With SCR-processed data, the multi-modal model has the best performance, it predicts the exact time when the object blocks the link with low absolute error (below 1 time instance (100 ms) for all prediction interval). Multi-modal model without SCR pre-processing and the model using SCR-processed LiDAR data are also able to predict the exact time instance with low error, below 1.1 time instances (110 ms) for all prediction intervals.

\noindent\textbf{Robustness Analysis:} The performance of the models that are trained in site 0 and tested in sites 1 and 2 are shown in \figref{fig:p2}. Similar to the performance in \textbf{Problem 1}, the performance of the model using wireless data doesn't change when the site changes. The models using LiDAR data without SCR processing perform poorly while the performance of models using LiDAR data with SCR processing are good and robust. Compared to the performance of the multi-modal model in \figref{fig:p2}, the performance degrades a little bit; the mean absolute error when the prediction interval is 10 time instances (1 s) drops to around 1.4 (140 ms) in site 1 and 1.3 (130 ms) in site 2. 

\begin{table}[]
\centering
\caption{Mean absolute error between the target (the exact time instance where the blockage happens) and the prediction of this blockage time instance when prediction time interval = 5 (500 ms) using ML method.}
\begin{tabular}{cccc}
\hline \hline
\multirow{2}{*}{Data modality} & \multicolumn{3}{c}{Problem 2 (time instance)} \\ 
                               & site 0   & site 1   & site 2  \\ \hline
mmWave                         & 0.65      & 0.64      & 0.62     \\
mmWave + LiDAR                 & 0.55      & 0.96      & 1.2     \\
mmWave + LiDAR SCR             & 0.51      & 0.64      & 0.62   \\ \hline \hline
\end{tabular}

\label{tab:p2}
\end{table}

\noindent\textbf{Summary:} We conclude that i) the model using wireless data is robust but it has a good performance for only the first 4 time instances (400 ms). ii) The SCR processing improves the performance when the models are trained in site 0 and tested in sites 1 and 2. It doesn't improve the performance dramatically when the models are trained and tested at the same site. iii) The multi-modal model using SCR-processed LiDAR data is the best for all three sites. \tabref{tab:p2} summarizes the performance of the model using mmWave data, mmWave, and LiDAR data with or without SCR in three locations when the prediction time interval is 5 time instances (500 ms). With SCR processing, the performance improves up to 0.6 time instance (60 ms) error compared to the performance without SCR processing.

We also conclude that ML models perform very well for this problem. The ML method learns the error from the last epoch and improves the model in every epoch. However, in the proposed baseline LiDAR method, the LS estimation only predicts the time instance without any ``learning'' process and has weaker performance compared to the ML method. Thus ML-based methods are the better choice for this problem.

\subsection{Blockage Severity Level Prediction (\textbf{Problem 3})} \label{subsec:result_p3}
\noindent\textbf{Baseline Method:}

\begin{figure}[t]
	\centering
	\begin{subfigure}[]{\linewidth}
		\centering
		\includegraphics[width=1\linewidth]{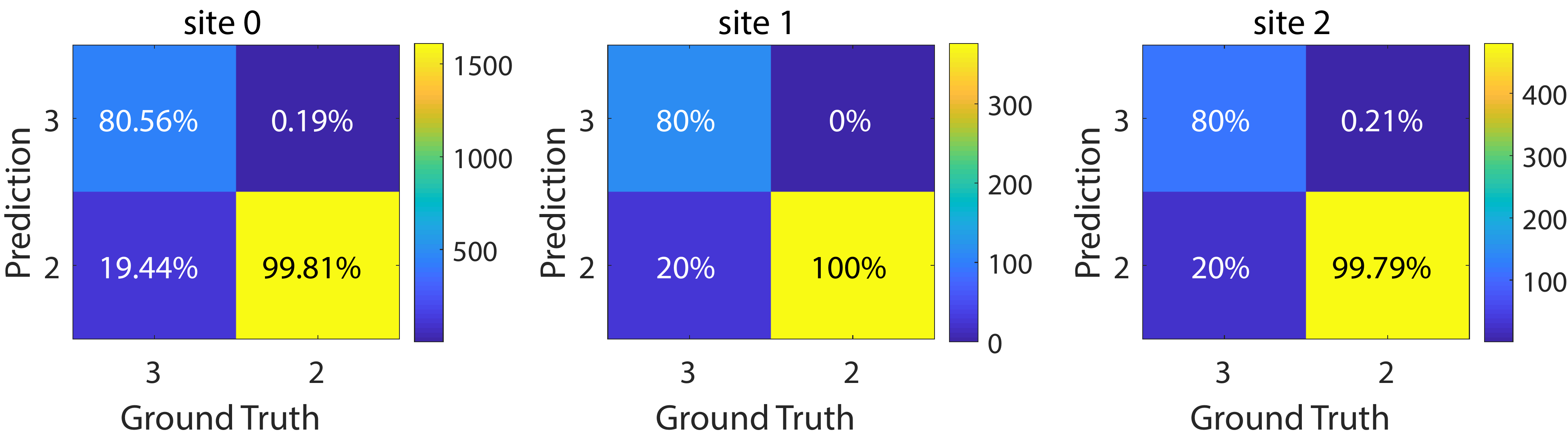}
		\caption{Blockage prediction with 2 severity levels}
		\label{fig:p3_trad_2cls}
	\end{subfigure}
	\begin{subfigure}[]{\linewidth}
		\centering
		\includegraphics[width=1\linewidth]{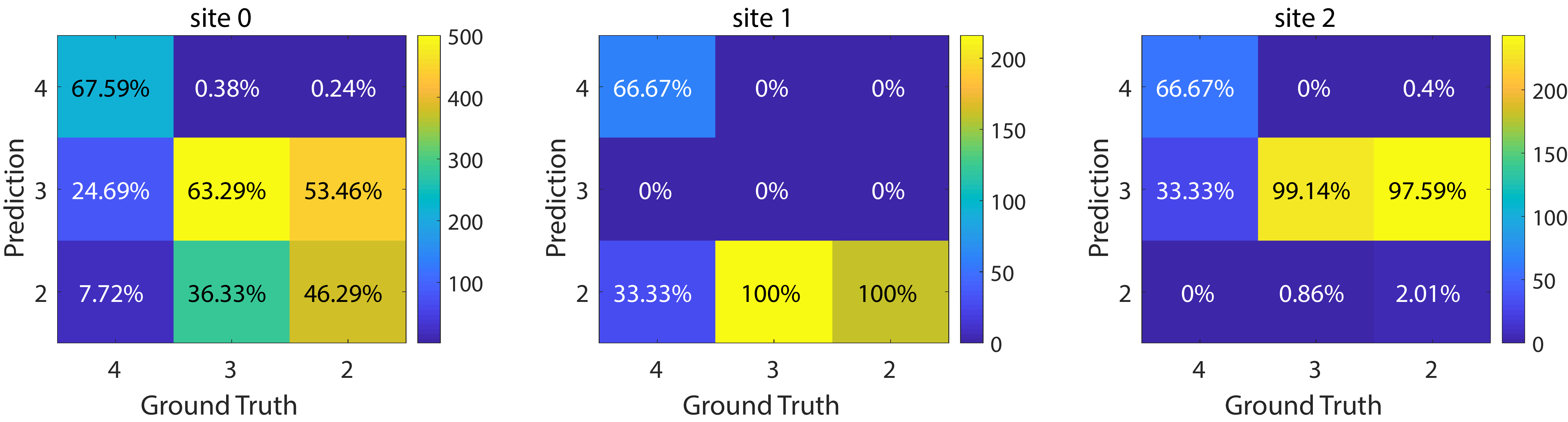}
		\caption{Blockage prediction with 3 severity levels}
		\label{fig:p3_trad_3cls}
	\end{subfigure}
	\caption{Confusion matrices for the severity level prediction with the baseline LiDAR method using the LiDAR SCR  dataset for prediction time interval = 2, the threshold is selected based on the data collected in site 0 and is tested for the data collected at all three locations. Subfigure (a) illustrates the case where we consider the objects in the classes with severity levels 2 and 3 and Subfigure (b) illustrates the case when considering the objects in the classes with severity levels 2, 3, and 4.}
	\label{fig:p3_trad}
\end{figure}

We follow the severity level classification in our previous work \cite{mmwave_journal}, for the case $N_\mathrm{class} = 3$, humans and bicycles fall in severity level 1, and campus carts, sedans, SUVs, pickup trucks, vans, and box trucks fall in severity level 2, large shuttles such as school shuttle, commercial trucks are in severity level 3. For the case $N_\mathrm{class} = 4$, humans and bicycles fall in severity level 1 and campus carts, sedans, and SUVs fall in severity level 2, middle size vehicles such as pickup trucks, vans, and box trucks are in severity level 3, and large shuttles are in severity level 4. \textbf{We placed the testbed close to the road, and there is no severity level 1 objects in our dataset, so for \textbf{Problem 3} in this paper, we only focus on vehicles (severity level 2, 3, 4).} The performance of predicting the blockage severity using the threshold-based method is shown by the confusion matrix in \figref{fig:p3_trad}. The vertical label is the prediction class and the horizontal label is the ground truth class. We present results for future prediction interval of $t = 2$ time instances (200 ms) for the case $N_\mathrm{class} = 3$. From the confusion matrix, we see that the performance is good for all three locations. We can predict the severity level 3 with 80\% accuracy and severity level 2 with 99\% or higher accuracy. When SCR is applied, most of the static clusters are removed, and the remaining points correspond to moving objects. The objects with high severity levels are represented with more points and objects with low severity are represented with fewer points. So setting a proper threshold helps distinguish the severity levels.

To better understand the blockage prediction problem, we take a step further. Since the threshold-based method can predict the case $N_\mathrm{class} = 3$ with high accuracy, we extend it to the case when $N_\mathrm{class} = 4$ \cite{mmwave_journal}. The result of the case $N_\mathrm{class} = 4$ prediction problem is shown in \figref{fig:p3_trad_3cls}. It has the best performance in site 0 since the threshold selection is based on the data collected in site 0. The performance is not so good in sites 1 and 2. For data collected in site 1, this method tends to predict either severity level 4 or 2. For data collected in site 2, the method predicts objects in severity level 4 or 3. Based on these results, we conclude that the threshold method works well for $N_\mathrm{class} = 3$ case but it fails for $N_\mathrm{class} = 4$.

\begin{figure}[t]
	\centering
	\begin{subfigure}[]{\linewidth}
		\centering
		\includegraphics[width=1\linewidth]{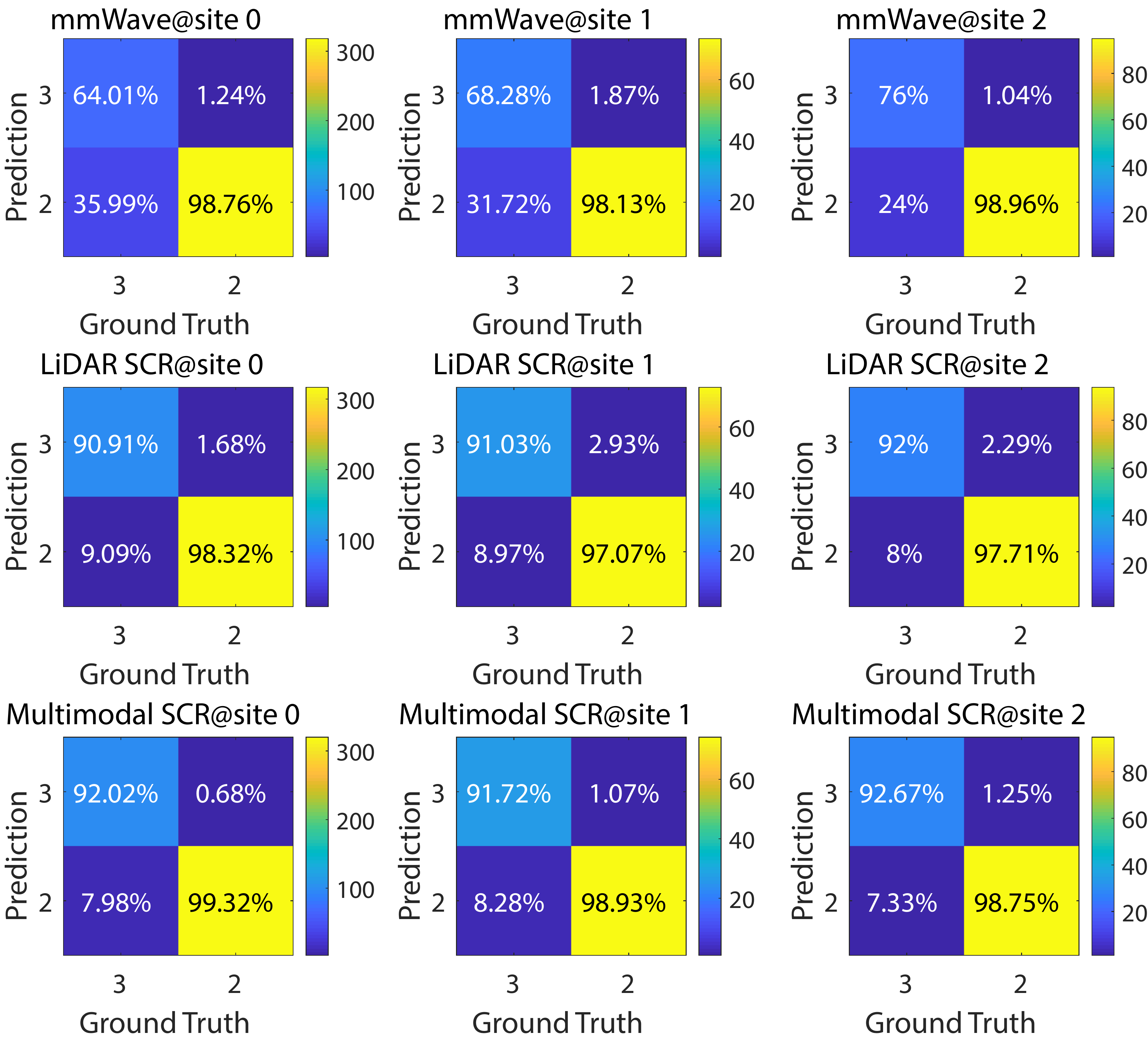}
		\caption{Blockage prediction with 2 severity levels}
		\label{fig:p3_ml_2cls}
	\end{subfigure}
	\begin{subfigure}[]{\linewidth}
		\centering
		\includegraphics[width=1\linewidth]{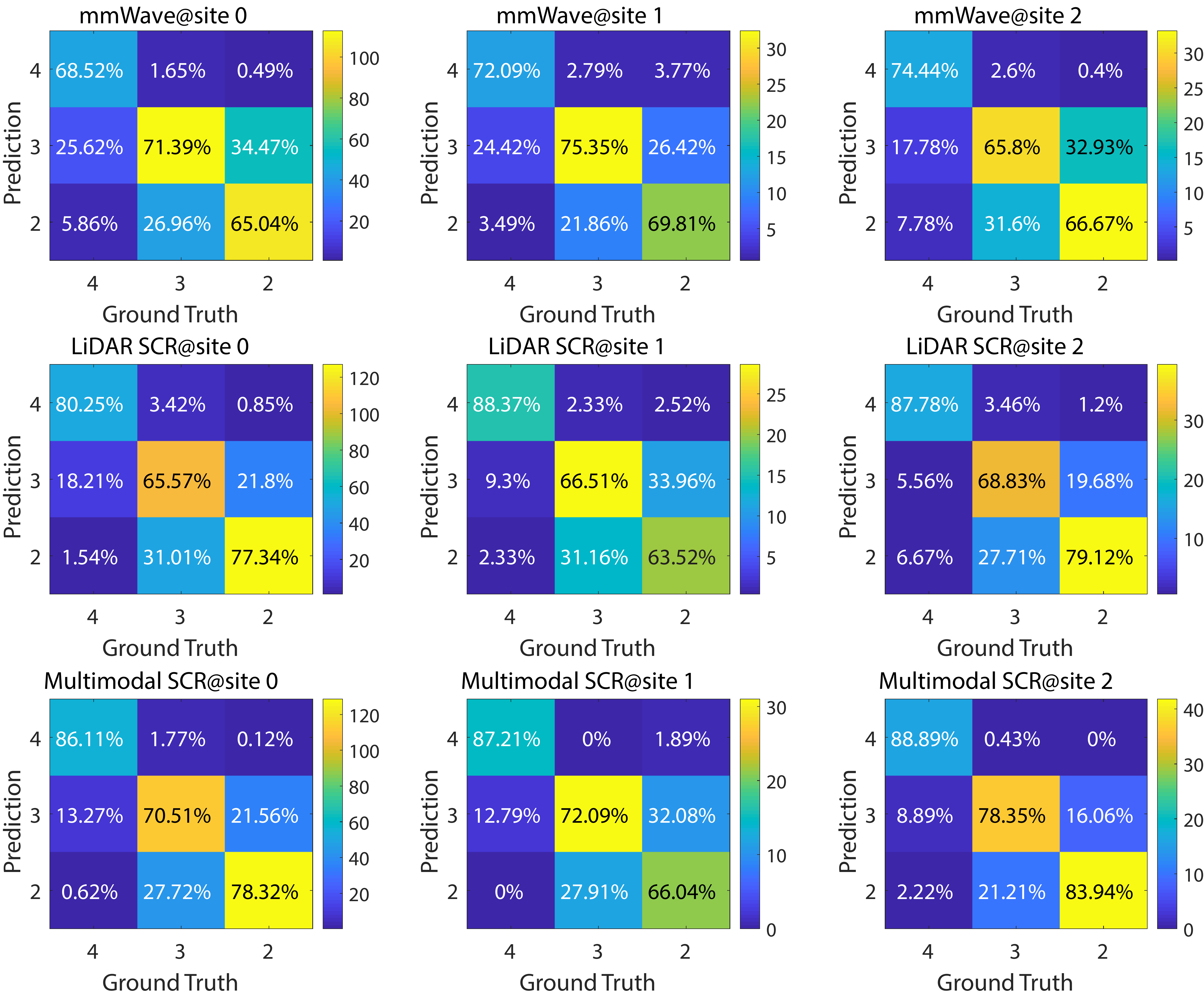}
		\caption{Blockage prediction with 3 severity levels}
		\label{fig:p3_ml_3cls}
	\end{subfigure}
	\caption{Confusion matrices for severity level prediction with the ML method using multimodal, LiDAR, or mmWave dataset for prediction time interval = 2 (200 ms). The model is trained using data collected in site 0 and is tested using data collected at all three locations. Subfigure (a) illustrates the case where we consider the objects in the classes with severity levels 2 and 3 and Subfigure (b) illustrates the case when considering the objects in the classes with severity levels 2, 3, and 4.}
	\vspace{-0.1cm}
	\label{fig:p3_ml}
\end{figure}

\noindent\textbf{ML Based Method:} The confusion matrix for both $N_\mathrm{class} = 3$ and $N_\mathrm{class} = 4$ cases prediction is shown in \figref{fig:p3_ml}; the prediction interval is 2 time instances (200 ms). \textbf{We only show the performance with SCR processing, since, without SCR processing, our model is not able to classify the severity level.} Similar to the proposed baseline LiDAR method, we train our model using data collected in site 0 and test it using the data collected in all three locations. The confusion matrices using data collected in site 0, site 1, and site 2 are presented in the left, middle and right panels, respectively. In both \figref{fig:p3_ml_2cls} and \figref{fig:p3_ml_3cls}, the confusion matrices in the top row use mmWave data to train and test our model, the matrices in the second and third row use LiDAR SCR data and multi-modal SCR data, respectively. For $N_\mathrm{class} = 4$ case shown in \figref{fig:p3_ml_3cls}, our model is able to predict the severity level with high accuracy using multimodal SCR and LiDAR SCR data collected in all three locations. The performance of wireless data is worse; the model can predict $N_\mathrm{class} = 3$ case but not the $N_\mathrm{class} = 4$ case. 

For $N_\mathrm{class} = 4$ case, shown in \figref{fig:p3_ml_3cls}, the performance using data collected in different locations are similar. The performance using multimodal SCR data is the best with above $86\%$ prediction accuracy for severity level 4, $70\%$ accuracy for severity level 3 and above $66\%$ accuracy for severity level 2. The performance using LiDAR SCR data is next followed by wireless data (weakest). 

\noindent\textbf{Summary:} We conclude that for both $N_\mathrm{class} = 3$ and $N_\mathrm{class} = 4$ cases, using multimodal SCR data gives better performance. The model is robust since the performance is consistent when the location of data collection changes. The performance using the proposed baseline LiDAR method and ML method differs for severity level 2 and severity level 3 cases. Since the threshold is selected based on the data collected in site 0, once the environment changes and SCR processing cannot ideally eliminate all the static cluster noise, the proposed baseline LiDAR method is not able to classify severity levels 2 and 3 with a fixed threshold. Thus, for severity level 3 case, which is a complex prediction problem, the ML method performs better.

\subsection{Blockage Direction Prediction (\textbf{Problem 4})} \label{subsec:result_p4}
\begin{figure*}[t]
\centering
\includegraphics[width=\linewidth]{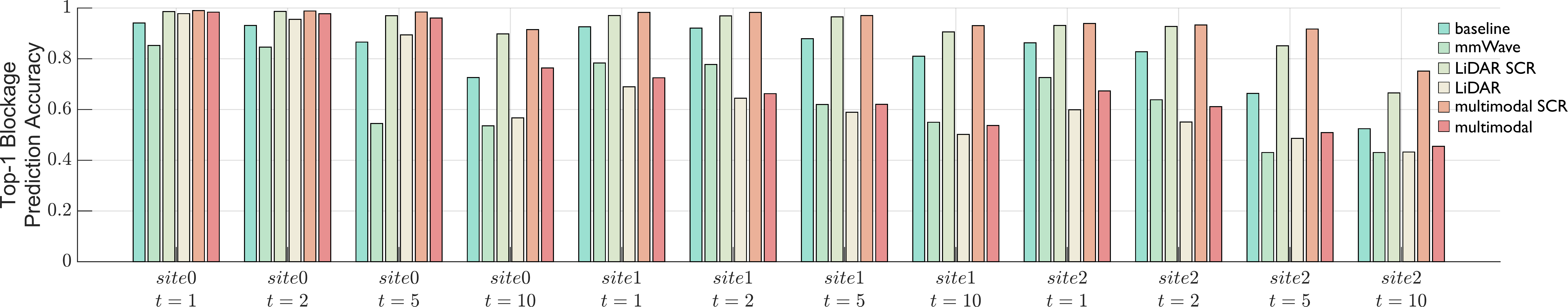}
\caption{Top-1 blockage direction prediction accuracy for prediction time interval 1, 2, 5, 10 time instances (duration of 1 time instance is 100 ms) using proposed baseline LiDAR method and ML-based method by inputting data collected at all three locations.}
\label{fig:p4}
\end{figure*}

The blockage direction prediction performance using the proposed baseline LiDAR method and ML-based method is shown in \figref{fig:p4}. We choose prediction time instance 1, 2, 5, and 10 and show the performance using the data collected at all three locations. The two moving directions are either from left to right or right to left.

\noindent\textbf{Baseline Method:}
 When the prediction interval is 1 time instance, we can predict the motion direction using the baseline method with 94\% accuracy using data in site 0; 93\% accuracy in site 1 is 93\% and 86\% in site 2. The accuracy decreases as the prediction interval increases. The performance in sites 0 and 1 is better than site 2 since the data in site 2 has more static cluster noise, which makes it harder to predict the moving direction.

\noindent\textbf{ML Based Method:} The results for blockage moving direction prediction are shown in \figref{fig:p4}. We plot Top-1 moving blockage direction prediction accuracy as a function of the future prediction interval using multimodal data, LiDAR and wireless data. For multimodal data and LiDAR data, we use both original and SCR-processed data. The results of data in sites 0, 1, and 2 all are shown in \figref{fig:p4}. The performance of the multimodal model using fused wireless and LiDAR data can achieve above 93\% accuracy for 1, 2, 5 time instances in the same site, but fails for correct prediction in other sites. While the multimodal model using fused wireless and LiDAR data achieves 95\% accuracy for the first time instance in sites 0 and 1; however the accuracy degrades as the prediction interval increases. The model can still achieve accuracy above 90\% using data in sites 0 and 1 and above 70\% in site 2. 

When using original multimodal or LiDAR data, the model achieves good accuracy in site 0 but the performance degrades dramatically in site 1 and site 2. Since LiDAR data includes the information of the moving blockages and surrounding environment, our model performs well when trained and tested in the same environment but is not able to perform well when the environment changes. Comparing the performance between original data and SCR processed data in site 1 and site 2, the SCR processing improves the robustness of our models. The model using wireless data achieves about 80\% accuracy in predicting the direction in the first time instance, but the performance decreases sharply after 3rd time instance. 

\begin{table}[]
\centering
\caption{The blockage motion direction accuracy when prediction time interval = 5 (500 ms).}
\begin{tabular}{cccc}
\hline \hline
\multirow{2}{*}{Data modality} & \multicolumn{3}{c}{Problem 4} \\ 
                               & site 0   & site 1   & site 2  \\ \hline
mmWave                         & 55\%      & 63\%      & 44\%     \\
mmWave + LiDAR                 & 96\%      & 63\%      & 69\%     \\
mmWave + LiDAR SCR             & 98\%      & 97\%      & 91\%   \\ \hline \hline
\end{tabular}

\label{tab:p4}
\end{table}

\noindent\textbf{Summary:} We conclude that the multimodal model has the best performance using fused LiDAR SCR and wireless data and it is robust to changes in environment. The model using fused data or LiDAR data without SCR processing performs well when trained and tested in the same environment, but the model does not perform well when the environment changes. The performance of the model using wireless data is similar when the surrounding environment changes, but the performance is not as good as LiDAR SCR data or fused LiDAR SCR/wireless data. \tabref{tab:p4} summarizes the performance of our model using mmWave data, mmWave, and LiDAR data with or without SCR in three locations when the prediction time interval is 5 time instances (500 ms). With SCR processing, the performance improves up to 44\% compared to the performance without SCR processing. By comparing the baseline method and ML-based method, the accuracy of the ML method is better in all three sites. Although the performance of both methods degrade using data collected in site 2, using the ML method still achieves high prediction accuracy. Thus, the ML method works better than the proposed baseline LiDAR method.

\section{Conclusion}

In this paper, we explored the potential of leveraging both in-band mmWave and LiDAR sensory data to proactively predict dynamic blockages in mmWave systems. This enables the network to make proactive decisions, e.g., helping basestations make decisions on hand-off and beam switching. We formulated the LiDAR and the fused LiDAR/wireless-aided blockage prediction problems and developed baseline and efficient ML models for these tasks. To validate the feasibility of the proposed approaches, we constructed a large-scale real-world mmWave-and-LiDAR dataset. Then, we designed a LiDAR data denoising (static cluster removal) algorithm that can enhance the data quality obtained from low-cost LiDAR sensors. Evaluating our developed solutions on this real-world dataset yields the following takeaways: 
\begin{itemize}
	\item For predicting future moving blockages that are within a short window (200 ms), using the mmWave pre-blockage signature approach in \cite{mmwave_journal} might be sufficient to achieve high prediction accuracy ($>$90\%). 
	\item For predicting moving blockages that are further in the future (up to 1s before the occurrence), our LiDAR-aided blockage prediction approach achieves more than 80\% top-1 accuracy with SCR processed data. 
	\item Applying static cluster removal/denoising processing can significantly improve the prediction accuracy, especially when low-cost LiDAR sensors are used. 
	\item In terms of 3GPP 5G NR latency, the proposed proactive blockage prediction approaches can achieve 10x improvement for the hand-off/beam switching tasks. 
	\item Our models are able to classify the shape/severity of the future blockages with more than 90\% accuracy and classify the direction of the moving blockages with more than 90\% accuracy with SCR processed data. 
	\item Our proposed blockage prediction approaches are shown to generalize well for unseen data collected from different sites (than those used in the training). 
\end{itemize}

For the future work, it is interesting to extend the proposed blockage prediction approaches to fuse more modalities. It is also important to extend these approaches to scenarios with distributed sensing at both the infrastructure and user devices.

\balance

\end{document}

%% file: input.tex
\usepackage{amsfonts}
\usepackage{times}
\usepackage{graphicx}
\usepackage{latexsym}
\usepackage{dsfont}
\usepackage{amssymb}
\usepackage{amsmath}
\usepackage{cite}
\usepackage{verbatim}

\newcommand{\figref}[1]{{Fig.}~\ref{#1}}
\newcommand{\tabref}[1]{{Table}~\ref{#1}}

\def\bb0{{\mathbb{0}}}

\def\ba{{\mathbf{a}}}
\def\bb{{\mathbf{b}}}

\def\bff{{\mathbf{f}}}

\def\bh{{\mathbf{h}}}

\def\bm{{\mathbf{m}}}

\def\b0{{\mathbf{0}}}

\def\bA{{\mathbf{A}}}
\def\bB{{\mathbf{B}}}

\def\bI{{\mathbf{I}}}

\def\bR{{\mathbf{R}}}

\def\bbE{{\mathbb{E}}}

\def\cA{\mathcal{A}}

\def\cN{\mathcal{N}}

\def\sf0{{\mathsf{0}}}

